\documentclass[letterpaper,twocolumn,10pt]{article}
\usepackage{usenix2019_v3}

\usepackage{enumitem}
\setlist{noitemsep,topsep=0pt,parsep=0pt,partopsep=0pt}
\setitemize{leftmargin=4mm}
\setenumerate{leftmargin=4mm}

\usepackage{enumitem}
\usepackage{amsmath}

\usepackage{amssymb}
\usepackage{xspace}
\usepackage{breakurl}
\usepackage{url}
\usepackage{multirow}
\usepackage{pifont}
\usepackage{caption}
\usepackage{subcaption}
\usepackage{booktabs}
\usepackage{algorithm}
\usepackage[noEnd=true, commentColor=gray]{algpseudocodex}
\tikzset{algpxIndentLine/.style={draw=black,thin}}
\errorcontextlines\maxdimen
\newcommand{\paraspace}{\vspace{0.01in}}
\newcommand{\parab}[1]{\paraspace\noindent{\bf #1} }

\newcommand{\etc}{etc.\xspace}
\newcommand{\ie}{\textit{i.e.}\xspace}
\newcommand{\eg}{e.g.\xspace}

\newcommand{\codefont}[1]{\texttt{#1}}
\newcommand{\sys}{\codefont{ConfigTuner}\xspace}

\usepackage{tikz}

\usepackage{makecell}

\usepackage[most]{tcolorbox}

\usepackage{booktabs}
\usepackage{tabularx}
\usepackage{array}
\usepackage{colortbl} 
\definecolor{Gray}{gray}{0.85}
\newcommand{\violet}{\textcolor{green!80!black}}

\begin{document}

\date{}

\title{Analyzing Communication Predictability in LLM Training}


\author{
\rm Wenxue Li$^{1}$~~~
\rm Xiangzhou Liu$^{1}$~~~
\rm Yuxuan Li$^{1}$~~~
\rm Yilun Jin$^{1}$~~~
\rm Zhenghang Ren$^{1}$~~~
\rm Xudong Liao$^{1}$~~~\\
\rm Han Tian$^{1}$~~~
\rm Bo Ren$^{1}$~~~
\rm Zhizhen Zhong$^{2}$~~~
\rm Guyue Liu$^{3}$~~~
\rm Ying Zhang$^{4}$~~~
\rm Kai Chen$^{1}$~~~\\
\\
$^1$Hong Kong University of Science and Technology~~~
$^2$Massachusetts Institute of Technology~~~\\
$^3$Peking University~~~
$^4$Meta~~~
}

\maketitle
\begin{abstract}
Effective communication is essential in distributed training, with predictability being one of its most significant characteristics. However, existing studies primarily focus on exploiting predictability through online profiling for runtime optimization, without a systematic understanding of it. In this work, we aim to systematically formulate communication predictability in distributed training, particularly in Large Language Models (LLMs) that utilize hybrid parallelism. Our analysis focuses on both traffic patterns and communication overhead. Specifically, we investigate predictable traffic patterns in typical LLMs and evaluate how various factors influence GPU utilization and effective bandwidth (two critical variables affecting communication overhead). Furthermore, we develop an analytical formulation to estimate communication overhead in LLM training, which is validated with high accuracy against empirical data. Leveraging this formulation, we propose a configuration tuning tool, \sys, to optimize training performance. Compared to Megatron-LM, the training configurations optimized by \sys demonstrate up to a 1.36$\times$ increase in throughput. Compared to Alpa, \sys generates the same configuration suggestion while significantly reducing the search complexity.
\end{abstract}

\section{Introduction}
Deep neural networks (DNNs) are increasingly adopted as fundamental building blocks in various modern services, such as language translation, autonomous driving, and chatbots~\cite{brown2020language, touvron2023llama, zhao2023survey}. Due to the limited capabilities of individual accelerators (\ie, GPUs) and the high computational demands of DNN training, it is typically distributed across hundreds or even thousands of accelerators~\cite{wang2020domain, zeng2024accelerating, hu2024characterization, jiang2024megascale, narayanan2019pipedream, narayanan2021efficient}. These accelerators work collaboratively using various parallelization strategies and rely heavily on communication to efficiently complete tasks. As a result, effective communication is crucial in distributed training, and numerous studies have explored various optimization techniques to address this challenge. These techniques include compressing transmitted traffic~\cite{li2024thc, bai2021gradient, wang2023hi, fei2021efficient} and enhancing the overlap between communication and computation~\cite{peng2019generic, ma2022autobyte, jayarajan2019priority}.


A deep understanding of communication characteristics is crucial for optimizing the communication process. Unlike traditional datacenter applications~\cite{benson2010understanding, roy2015inside, benson2010network}, distributed training exhibits unique communication characteristics, with predictability being the most significant. However, existing research on communication predictability in distributed training remains limited. On one hand, most studies focus on leveraging predictability to enhance system performance~\cite{xiao2018gandiva, rajasekaran2023cassini, sivathanu2019astra, wang2023topoopt, wang2023build}. These approaches typically rely on online profiling to gather statistics for future decisions in subsequent iterations, but they generally lack a deep understanding of predictability itself. On the other hand, existing research only touches on the repetitive nature of communication patterns across iterations, without providing an in-depth analysis of predictability~\cite{liu2022modeling, zhu2020daydream, wang2023build, zhang2020network, hu2024characterization, gangidi2024rdma, qian2024alibaba}.


In this paper, we aim to provide a comprehensive analysis of communication predictability in distributed training, with a particular focus on Large Language Models (LLMs) that employ various parallelism strategies. Our analysis covers representative LLM architectures, including both densely-activated Generative Pre-trained Transformer (GPT) models that use hybrid parallelism and sparsely-activated Mixture of Experts (MoE) models with expert parallelism. Specifically, we examine two key aspects of communication in LLM training: traffic patterns and communication overhead. Traffic patterns refer to the high-level spatial and temporal characteristics of communication, such as traffic volume and communication matrix, while communication overhead represents the time and ratio spent on communication during the overall training.

First, we comprehensively analyze the traffic patterns during the training of LLMs. Specifically, we empirically validate the predictability and semi-predictability of these patterns. Our observations show that the traffic patterns of GPT models can be accurately predicted based on the model architecture and parallelism configuration, without the need for online profiling. In contrast, the traffic of MoE models is dynamic but exhibits semi-predictable characteristics ($\S$\ref{pattern}).

Next, we develop an analytical formulation to formulate the communication overhead by modeling the various communication phases and the overall iteration time during LLM training. Our formulation has been validated with high accuracy against empirical results ($\S$\ref{modeling}). Two critical variables influence communication overhead in our analytical formulation: the effective bandwidth during the communication phase and GPU utilization rate during the computation phase. To understand their impact, we conduct an extensive evaluation of various factors. We observe that GPU utilization rate remains relatively stable when the per-GPU model size and micro-batch size are fixed. Additionally, the maximum potential of effective bandwidth is constrained by the capacity of GPU interconnects, with factors such as message size, collective scale, and synchronization overhead lowering the practical effective bandwidth than the theoretical upper limit ($\S$\ref{benchmark}).

Ultimately, we present a configuration tuning tool called \sys to optimize the LLM training performance ($\S$\ref{optimization}), demonstrating the practical utility of the analytical formulation in $\S$\ref{modeling} and empirial analysis in $\S$\ref{benchmark}. \sys only requires minimal profiling data, while other configuration tuning tools, such as Alpa~\cite{zheng2022alpa}, typically require running multiple full iterations for each configuration. We demonstrate the effectiveness of \sys in three optimization scenarios, where the training configurations determined by \sys achieve up to a 1.36$\times$ increase in throughput compared to Megatron-LM. Moreover, \sys provides the same configuration suggestions as Alpa while significantly reducing search complexity.


This work presents several key findings and contributions:
\begin{itemize}[leftmargin=*]
	\item We systematically formulate and validate the predictability of LLM training, advancing the understanding from repetitive patterns to pre-execution determination.
	\item We thoroughly explore the predictable traffic patterns in the training of GPT models and uncover the semi-predictability in the training of MoE models.
	\item We propose an analytical formulation to precisely estimate the communication overhead for LLMs, offering a mathematic understanding of predictability. We also extensively benchmark two critical variables in our formulation. 	
	\item We develop a configuration tuning tool called \sys to optimize training performance by identifying the most efficient training configuration. \sys showcases the practical application of our analytical formulation.
\end{itemize}

\section{Background and Motivation}
We begin by presenting production data from our partner company to offer an overview of the communication overhead in real-world production environments ($\S$\ref{sec2:meta}). Next, we demonstrate insufficient explorations on communication predictability in LLM training and outline our objectives ($\S$\ref{sec2:goal}). Finally, we introduce the parallelism strategies and associated communication operations used in LLM training, which form the foundation for our subsequent analysis ($\S$\ref{sec2:back}).

\subsection{Communication Overhead in Production}\label{sec2:meta}
We analyze traffic traces from several typical DNN training jobs provided by our partner company. Specifically, we measure the iteration time and the percentage of communication time during iterations (\ie, communication ratio) of typical DNN training jobs, to provide a basic overview of the communication overhead in production environments.

\begin{figure}[h]
	\centering
	\begin{subfigure}[t]{0.9\linewidth}
		\includegraphics[width=\linewidth]{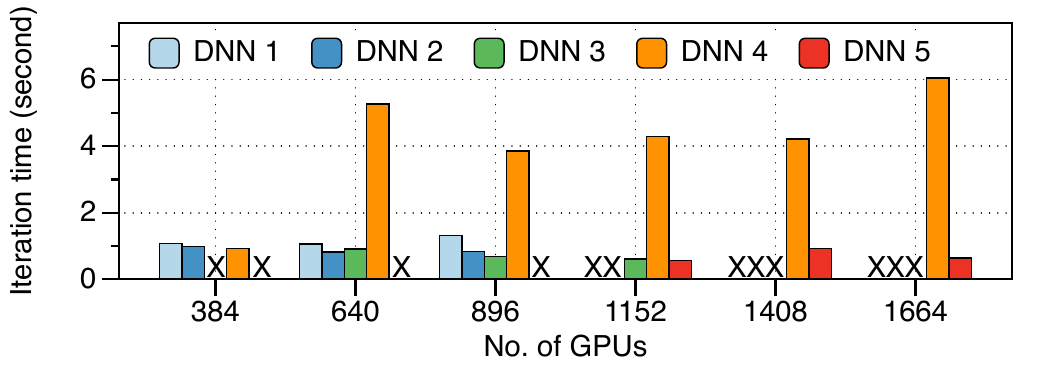}
		\caption{Iteration time of five DNN models.}
		\label{fig:meta:1}
	\end{subfigure}
	\begin{subfigure}[t]{0.9\linewidth}
		\includegraphics[width=\linewidth]{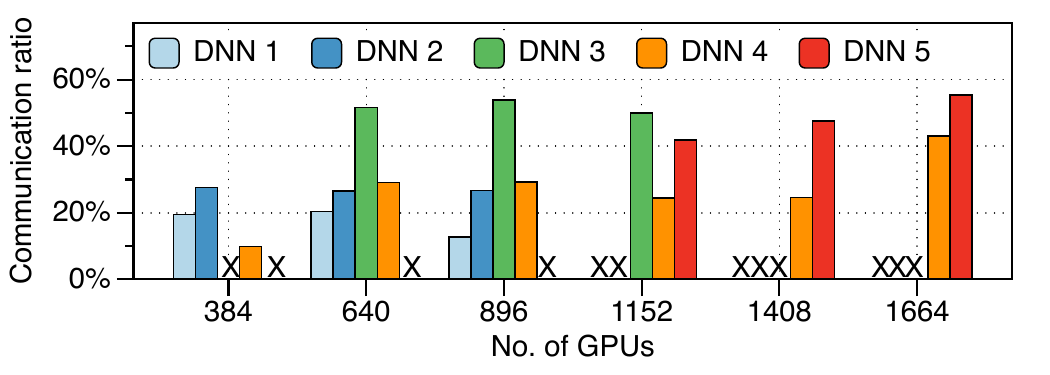}
		\caption{Communication ratio of five DNN models.}
		\label{fig:meta:2}
	\end{subfigure}
	\caption{Production measurements demonstrate high communication overhead in large-scale DNN training.}
	\label{fig:meta}
\end{figure}

Fig.~\ref{fig:meta} illustrates the iteration times and communication overhead during the training of five DNN models, using varying numbers of GPUs, ranging from 384 to 1664. Some models lack values under certain scales, denoted by the $\times$ symbol, and the exact names of DNN models are not shown for confidentiality reasons. The results demonstrate that communication overhead accounts for up to 20\%, 27\%, 53\%, 43\%, and 55\% during the training of these five models, respectively. These findings underscore the significant communication overhead experienced in production environments and highlight the critical need for in-depth analysis.

\subsection{Insufficiency of Existing Works}\label{sec2:goal}
Distinct from traditional datacenter applications~\cite{benson2010understanding, roy2015inside, benson2010network}, predictability is the most significant communication characteristic of distributed training. However, existing research on communication predictability in distributed training remains limited. Most studies focus on leveraging predictability to enhance system performance in areas such as communication scheduling~\cite{xiao2018gandiva, rajasekaran2023cassini, sivathanu2019astra,hashemi2019tictac,gu2019tiresias} and datacenter topology configuration~\cite{wang2023topoopt, wang2023build}, primarily using online profiling to gather runtime statistics for future decisions in subsequent iterations. Yet, these approaches often lack deep exploration of the predictability itself. Additionally, existing research typically only touches on the repetitive nature of communication patterns across iterations but falls short of offering an in-depth analysis of communication predictability~\cite{liu2022modeling, zhu2020daydream, wang2023build, zhang2020network, hu2024characterization, gangidi2024rdma, qian2024alibaba}.

In this paper, we aim to fill this gap by offering a comprehensive analysis of communication predictability in distributed training, particularly in LLMs that employ various parallelism strategies. We focus on two key aspects: traffic patterns and communication overhead, conducting an in-depth analysis supported by accurate mathematical modeling and empirical evaluations.



\subsection{Parallelism Strategies in LLM Training}\label{sec2:back}
In our paper, our focus is on the parallelism techniques in LLMs for two reasons. On one hand, the parallelism strategies employed in LLM training are among the most diverse of all DNNs. On the other hand, LLM training imposes extremely high computational demands; as a result, hundreds or even thousands of accelerators communicate intensively to complete tasks, emphasizing the critical role of communication~\cite{wang2020domain, zeng2024accelerating, hu2024characterization, jiang2024megascale, narayanan2019pipedream, narayanan2021efficient}. This subsection introduces the most widely used parallelism strategies in LLM training, which serve as the foundation for our subsequent analysis.

\begin{figure}[t]
	\centering
	\includegraphics[width=\linewidth]{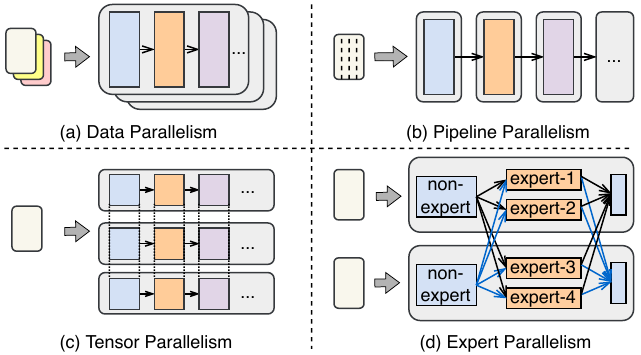}
	\caption{Four parallelism strategies used in distributed LLM training. Each color represents a different layer.}
	\label{fig:parallelism}
\end{figure}

\parab{Data parallelism (DP).} Data parallel training is widely adopted when a model can fit within a single accelerator. As shown in Fig.~\ref{fig:parallelism}\violet{a}, each accelerator holds a replica of the model and a subset of the data batch. During each iteration, the accelerator performs forward and backward computations locally, and all accelerators synchronize their gradients using one or several \codefont{AllReduce} operations. The AllReduce operation can be implemented using various techniques, such as parameter servers~\cite{li2014communication, jiang2020unified}, Ring/Tree-AllReduce~\cite{thakur2005optimization, nccl}, and hierarchical AllReduce~\cite{cho2019blueconnect}.

\parab{Pipeline parallelism (PP).} For large models that exceed the memory capacity of a single accelerator, the model must be divided across multiple accelerators using model parallelism. Pipeline parallelism divides the model layers "vertically", creating several pipeline stages~\cite{narayanan2019pipedream, huang2019gpipe}. Moreover, PP splits each data batch into many micro-batches and pipelines them through the stages. The illustration of PP is shown in Fig.~\ref{fig:parallelism}\violet{b}. The primary overheads of PP are the pipeline bubble time and the point-to-point (P2P) communication, \ie, \codefont{Send} or \codefont{Recv}, time between adjacent stages.

\parab{Tensor parallelism (TP) and sequence parallelism (SP).} As shown in Fig.~\ref{fig:parallelism}\violet{c}, tensor parallelism divides each layer "horizontally," with each accelerator responsible for a portion of the tensor computations of each layer. TP is commonly used in training GPT models and incurs four \codefont{AllReduce} operations per transformer layer (two for each attention and MLP block)~\cite{narayanan2021efficient}. Unlike DP, TP's AllReduce operations typically cannot overlap with computation. Another notable model parallelism approach is sequence parallelism, which was developed to mitigate the high memory consumption associated with TP, through re-sharding the activations after synchronization. SP can be considered a variant of TP, exhibiting similar traffic volume and communication time but with lower activation memory consumption~\cite{korthikanti2023reducing}. Given our focus on the analysis of communication, we solely consider TP in this work and assume SP to be equivalent to TP. 

\parab{Hybrid parallelism.} Distributed training for large GPT models commonly employs a combination of data, pipeline, and model parallelism (\ie, hybrid parallelism). Megatron-LM~\cite{narayanan2021efficient} terms its hybrid parallelism, which consists of PP, TP, and DP, as PTD Parallelism (PTD-P). In this paper, we also refer to PTD-P when discussing hybrid parallelism.

\parab{Expert parallelism (EP).} Expert parallelism is used for distributed training for sparsely-activated LLMs, such as GPT-MoE models. As shown in Fig.~\ref{fig:parallelism}\violet{d}, the GPT-MoE architecture typically replaces a dense MLP layer with an expert layer consisting of multiple experts, and uses EP to distribute these experts across multiple accelerators~\cite{fedus2022switch, shazeer2017outrageously}. EP requires an \codefont{AllToAll} operation before and after each expert layer to exchange intermediate results. 



\section{Analysis of Traffic Pattern}\label{pattern}
In this section, we explore the traffic patterns of distributed training in both spatial and temporal dimensions. Spatial patterns are characterized using traffic heatmaps, determined by two primary elements: communication matrix and traffic volume, identifying the pairs of GPUs that carry traffic and the amount of traffic on each GPU pair, respectively. Temporal patterns are illustrated by the momentary transmited traffic volume over time. We initiate our exploration with an in-depth analysis of predictability in densely-activated GPT models ($\S$\ref{predic}) and subsequently analyze the semi-predictability in sparsely-activated MoE models ($\S$\ref{moe}).

\subsection{Predictability of Densely-activated LLMs}\label{predic}
For the traffic heatmaps of densely-activated GPT models, we find that \textit{both the communication matrix and traffic volume are predictable}, with the primary determinants being the parallelism configuration and model architecture. Additionally, the variation in momentary transmited traffic volume during the training of GPT models exhibits \textit{a regular “On-Off” transmission feature}, characterized by predictable and repetitive durations for both the "On" and "Off" periods. 

\begin{figure}[t]
	\centering
  	\includegraphics[width=\linewidth]{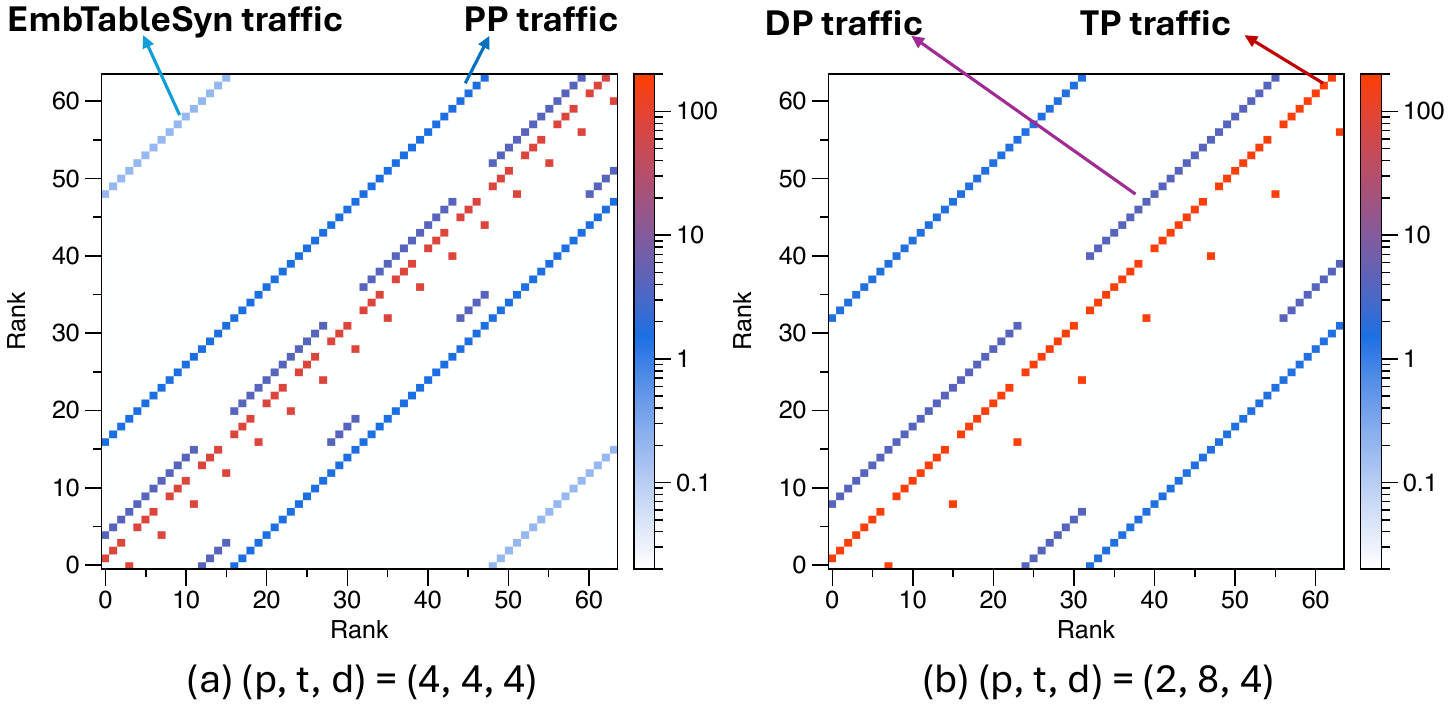}
	\caption{Traffic heatmaps of a GPT model with 39B parameters with two parallel configurations on 64 GPUs.}
	\label{fig:heatmap}
\end{figure}

\parab{Parallelism configuration determines the communication matrix.} It is widely known that the communication matrix of purely data parallel models forms a diagonal line, regardless of the specific model architectures~\cite{liu2022modeling, zhu2020daydream, wang2023build, zhang2020network}. For models utilizing hybrid parallelism, the communication matrix is more complex than in purely data parallel training but is still determined by the parallelism configuration. For example, we conduct training of a GPT model with 39 billion parameters across 64 H-series (Hopper architecture) GPUs and evaluate two parallelism configurations: (pipeline-, tensor-, data-parallel size) $(p,t,d)$ = $(4, 4, 4)$ and $(2, 8, 4)$. The corresponding traffic heatmaps are depicted in Fig.~\ref{fig:heatmap}. Note that the colorbar unit is in gigabytes (GB), the white squares indicate no traffic, and the red, purple, and dark blue squares represent TP, DP, and PP traffic, respectively.

\begin{figure}[t]
	\centering
	\includegraphics[width=0.9\linewidth]{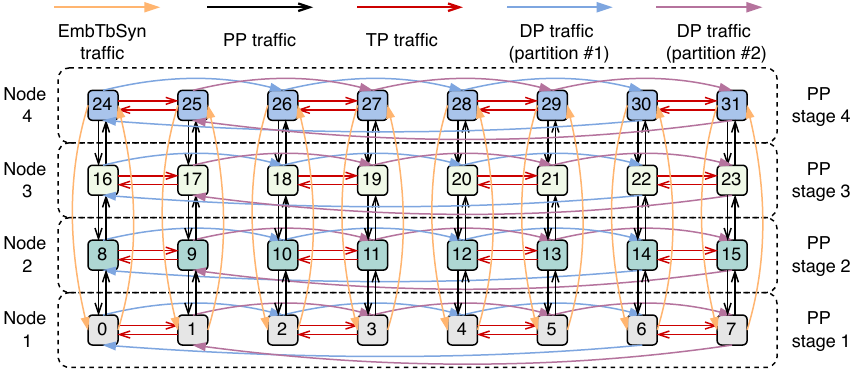}
	\caption{Communication matrix of a parallelism configuration $(p,t,d)=(4, 2, 4)$ on 32 GPUs (4 machines).}
	\label{fig:mapping}
\end{figure}

As results show, both the TP and DP traffic follow an AllReduce pattern structure. The PP traffic comprises two diagonal lines for transmitting activations during forward and backward phases. The light blue squares are for Embedding Table Synch (EmbTableSyn) traffic, occurring between the first and last PP stages to aggregate gradients of embedding tables. The two heatmaps are different due to different parallelism configurations. For example, Fig.~\ref{fig:heatmap}\violet{b} exhibits fewer TP groups each with more members and more inter-machine DP traffic. Note that the PP and EmbTbSyn traffic overlaps, resulting in only dark blue squares being visible in Fig.~\ref{fig:heatmap}\violet{b}.

\parab{Communication matrix is predictable.} Given a parallelism configuration, the communication matrix can be directly determined without actually running the model and conducting online profiling. Specifically, under typical hybrid parallelism, GPUs are organized into several PP stages, and within each stage, into several TP and DP groups. Each GPU belongs to one PP stage, one TP group, and one DP group, simultaneously. A logical parallelism configuration should be mapped to a physical hardware platform for real deployment. Assuming we adopt a mapping principle that tries to allocate TP and DP groups within machines and distribute PP stages across machines, we could depict the communication matrix before actually running the model. For example, Fig.~\ref{fig:mapping} illustrates the communication matrix for a logical strategy $(p,t,d) = (4, 2, 4)$ on 4 machines each with 8 GPUs. The $i_{th}$ GPU in stage $j$ transmits activations to the $i_{th}$ GPU in stage $j+1$ and $j-1$. Each GPU also participates in an AllReduce process for its corresponding TP and DP groups.

\begin{figure}[t]
	\centering
	\includegraphics[width=0.9\linewidth]{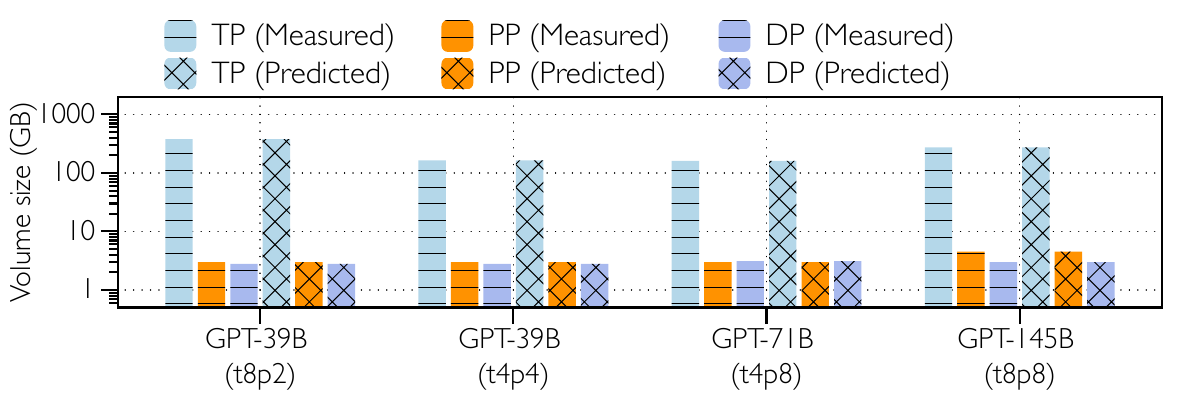}
	\caption{Traffic volume for three types of communication from one iteration under varying GPT model size.}
	\label{fig:volume}
\end{figure}

\parab{Model architecture determines traffic volume.} After establishing the parallelism configuration, it is the model's internal architecture that influences the traffic volume on GPU pairs. For example, in purely data parallel training, the total traffic volume is equal to $N \times \text{precision} \times \frac{2(d-1)}{d}$ (assuming Ring-AllReduce is adopted) where $N$ represents the number of parameters and $d$ denotes the data parallel size~\cite{zhang2020network, wang2020domain,liu2022modeling, zhu2020daydream}. Model architecture also impacts the traffic volume under hybrid parallelism. Fig.~\ref{fig:volume} shows the measured and predicted traffic volumes for different types of traffic in one training iteration of three GPT models ranging from 39 billion to 145 billion parameters. The results highlight varying traffic volumes based on model size, with TP traffic contributing more than 99\% of the total volume—a consistent trend across all our experiments. We exclude EmbTableSyn traffic from the analysis as it is negligible compared to the others.

\begin{figure}[t]
	\centering
	\includegraphics[width=0.99\linewidth]{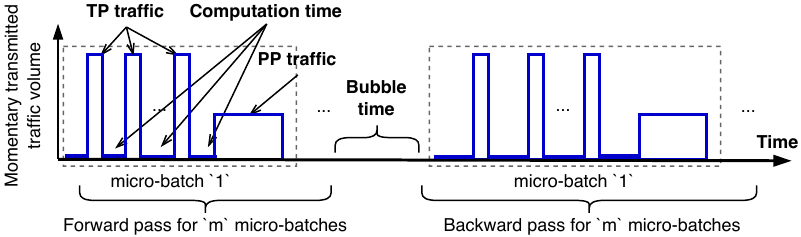}
	\caption{Illustration of the regular “On-Off” patterns with predictable and repetitive durations during one iteration.}
	\label{fig:onoff}
\end{figure}

\begin{figure*}[h]
	\centering
	\begin{subfigure}[t]{0.22\linewidth}
		\includegraphics[width=\linewidth]{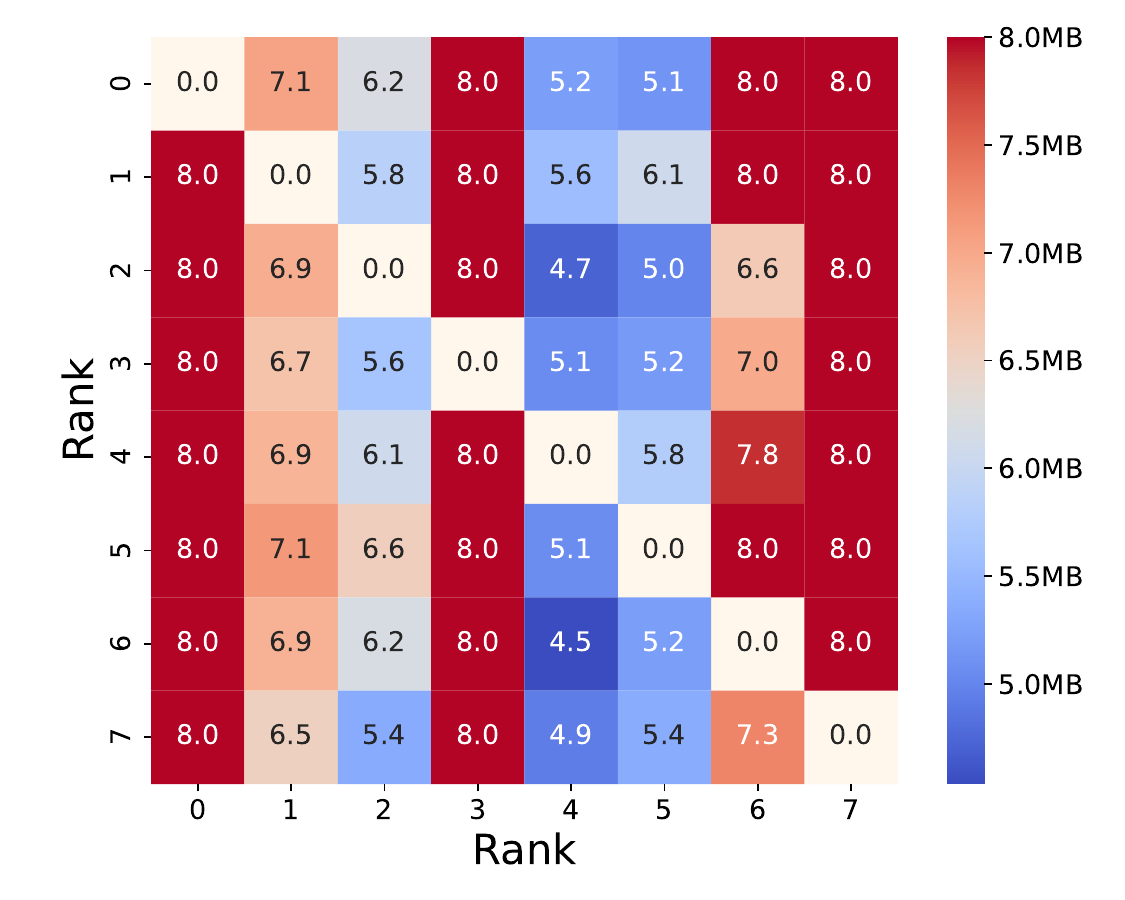}
		\caption{1st AllToAll in FW.}
		\label{fig:moe:fwbw:1}
	\end{subfigure}
	\begin{subfigure}[t]{0.22\linewidth}
		\includegraphics[width=\linewidth]{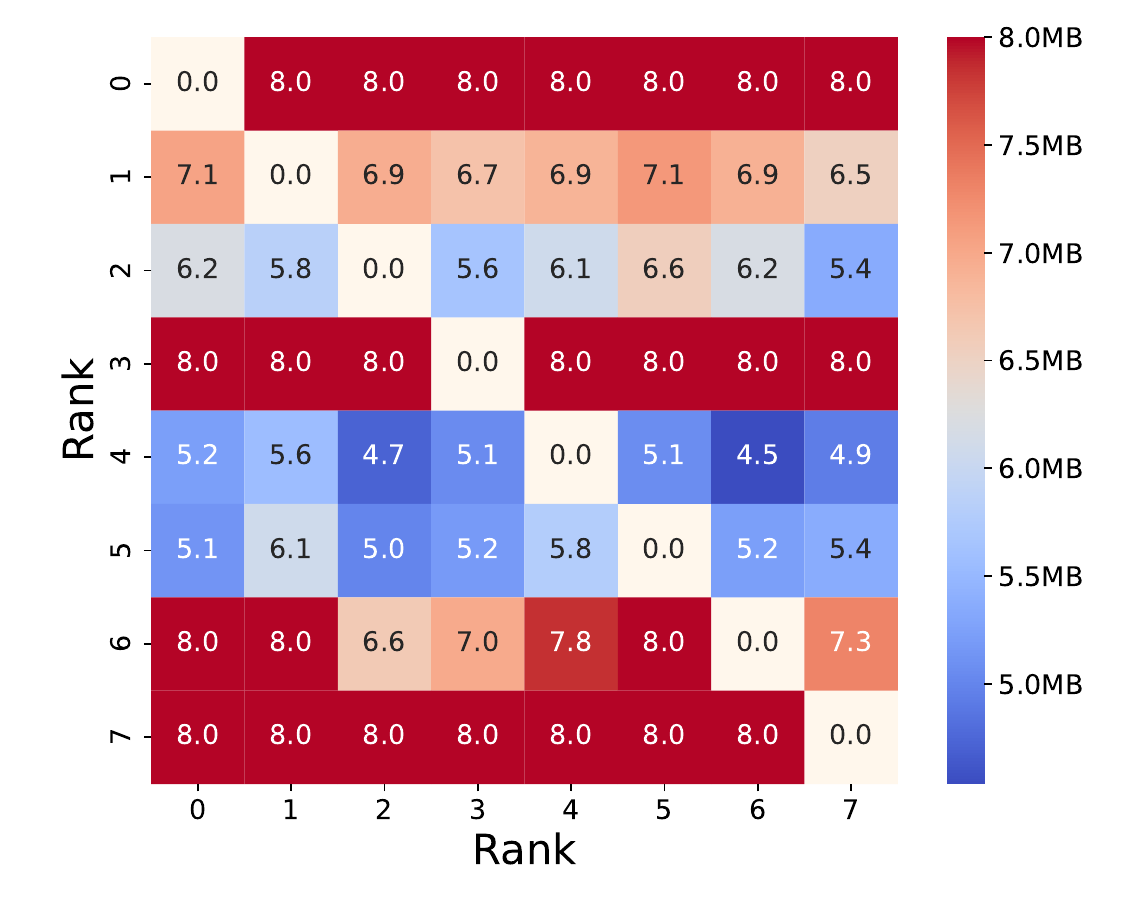}
		\caption{2nd AllToAll in FW.}
		\label{fig:moe:fwbw:2}
	\end{subfigure}
	\begin{subfigure}[t]{0.22\linewidth}
		\includegraphics[width=\linewidth]{figures/fw-iteration0_1}
		\caption{1st AllToAll in BW.}
		\label{fig:moe:fwbw:3}
	\end{subfigure}
	\begin{subfigure}[t]{0.22\linewidth}
		\includegraphics[width=\linewidth]{figures/fw-iteration0_2}
		\caption{2nd AllToAll in BW.}
		\label{fig:moe:fwbw:4}
	\end{subfigure}

	\caption{The four AllToAll communications at each layer are either identical or transposed.}
	\label{fig:moe:fwbw}
\end{figure*}



\parab{Traffic volume is computable.} Given the model architecture and parallelism configuration, the traffic volume can be precisely calculated without actually running the model. For instance, given a determined model architecture $(l, h, s, g, b, m)$ and parallelism configuration $(p,t,d)$ for GPT models (notations explained in Table~\ref{table:notation}), we can precisely calculate the traffic volume (including the detailed number of messages and the size of each message) on each GPU pair and accurately align with the observed results (as shown in Fig.~\ref{fig:volume}). The specific formula for this calculation is detailed in $\S$\ref{modeling}.


\parab{Regular "On-Off" patterns.} We selected a GPU from the training of a GPT-145B model and measured its momentary transmitted traffic volume during one iteration. We observe that it exhibits a structured and regular “On-Off” communication pattern in temporal dimension. The “On” period involves traffic transmission, while the “Off” period is for computation. Specifically, in a typical hybrid parallel training with $m$ micro-batches, a GPU experiences $2m$ micro-batch communication blocks ($m$ for each forward and backward pass). Each block consists of multiple TP and one PP traffic instances. The micro-batch communication blocks are repetitive, with consistent TP and PP durations throughout the training process. We show a simplified illustration in Fig.~\ref{fig:onoff}. 

\parab{Takeaway \#1:} \textit{For GPT models, the traffic heatmaps in the spatial dimension can be accurately predicted based on the model architecture and parallel configuration, eliminating the need for online profiling. The momentary transmitted traffic volume in the temporal dimension exhibits regular “On-Off” patterns with repetitive communication blocks.}

\subsection{Semi-predictability of Sparse LLMs}\label{moe}
The MoE architecture is a prominent method for implementing sparsely-activated LLMs. Without the loss of generality, we use GPT-MoE as a representative model of MoE models\footnote{In this paper, when we refer to MoE models, we specifically mean GPT-MoE models unless otherwise stated. The traffic characteristics of other representative MoE variants are similar to those of GPT-MoE models. For example, DeepSeek~\cite{dai2024deepseekmoe} and QWen~\cite{yang2024qwen2} differ only in that they utilize shared experts, while all other aspects remain the same.}. This architecture typically includes non-expert attention layers and replaces a dense MLP layer with an expert layer composed of multiple \textit{experts} and a \textit{gate} network. Training large MoE models requires expert parallelism (EP) to distribute experts across multiple accelerators~\cite{fedus2022switch,shazeer2017outrageously}. As depicted in Fig.~\ref{fig:parallelism}\violet{d}, the gate network receives output tokens from the preceding non-expert layer and selects one or several experts with the highest relevance to be active for each token. These active experts may be located on different accelerators, resulting in \codefont{AllToAll} operations. Since expert selection is determined at runtime and varies depending on the input tokens, AllToAll communication is dynamic during MoE training.

\begin{figure}[t]
	\centering
	\includegraphics[width=\linewidth]{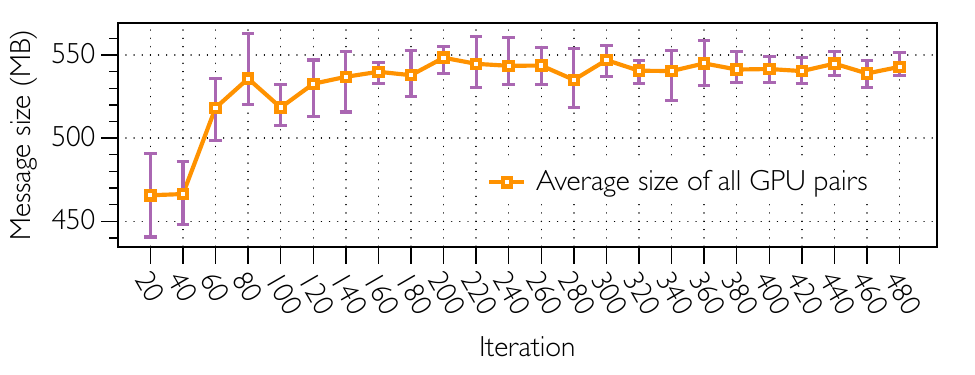}
	\caption{Average traffic volume and the variance across all GPU pairs during a 1.3B MoE model's training.}
	\label{fig:moe:adaptive}
\end{figure}

\parab{Predictability to some degree (semi-predictability).} Despite the inherently dynamic nature of AllToAll traffic, certain aspects exhibit predictability. 

First, for a given data batch, there are four AllToAll communications at one layer: two in the forward and two in the backward pass. \textit{These communications are either identical or transposed.} Specifically, the second AllToAll is the transpose of the first AllToAll in the forward pass. Additionally, the AllToAll communications in the backward pass are identical to those in the forward pass because the tokens must return to the previously selected experts for gradient calculation. To validate this, we measured the AllToAll communication heatmaps at the first expert layer during the training of the MoE-1.3B model and depicted the results of the four AllToAll communications in Fig.~\ref{fig:moe:fwbw}. The results demonstrate that the second AllToAll in both the forward and backward passes is the transpose of the first AllToAll in the forward pass, and the first AllToAll in the backward pass is identical to the first AllToAll in the forward pass. This consistency enables us to accurately predict the subsequent three AllToAll patterns based on the first AllToAll pattern in the forward pass.

Second, the gate network in MoE models is trained to achieve load balancing (achieve load balancing across experts)~\cite{rajbhandari2022deepspeed,fedus2022switch}, leading to \textit{increasing uniformity in AllToAll traffic patterns as training progresses}. To validate the phenomenon of increasing uniformity, we measured the average volume and variance of AllToAll traffic across all GPU pairs over the first 500 training iterations of the MoE-1.3B model\footnote{In detail, we first obtain an overall AllToAll communication heatmap during one iteration. For example, with 8 GPUs, there are 64 GPU pairs. We calculate the average traffic volume and variance across these 64 GPU pairs.}. As shown in Fig.~\ref{fig:moe:adaptive}, the average AllToAll volume stabilizes, and the traffic variance within each iteration consistently decreases as training progresses. This observation supports the hypothesis that the gate network distributes tokens more uniformly as training advances. 

Additionally, we observed an increase in the average AllToAll volume over time. This increase is attributed to the adoption of an adaptive top-2 gating algorithm~\cite{lepikhin2020gshard}, where each token can select one or two experts dynamically, thus causing fluctuating traffic volumes. A trend observed during our experiments is that as training progresses, tokens tend to select two experts instead of one expert with increasing possibility, thus increasing the AllToAll traffic volume.


\parab{Takeaway \#2:} \textit{The AllToAll traffic is dynamic due to the online decision-making from gate network, yet it exhibits certain semi-predictable patterns, such as the consistency in the four AllToAll communications and the increasing uniformity in expert selection as training progresses.}

\section{Analytical Formulation}\label{modeling}
In this section, we first develop an analytical formulation for estimating the communication overhead of GPT ($\S$\ref{formulation:gpt}) and MoE models ($\S$\ref{formulation:moe}). We then validate the accuracy of this formulation through experiments by comparing its predictions with actual measured results ($\S$\ref{formulation:vali}). Note that the training process of purely data-parallel models is a subset of that of GPT models; therefore, the formulation of GPT models also applies to purely data-parallel models.

\subsection{Modeling for GPT Models}\label{formulation:gpt}
All our analysis and experiments for GPT models leverage the hybrid-parallel training and mixed-precision training, \ie, utilizing 16-bit precision for model parameters, activations, and gradients. For the pipeline scheduling, we adopt the widely-used "1F1B" scheme~\cite{narayanan2019pipedream}. The used notations are illustrated in Table~\ref{table:notation}. We begin the analysis with a dissection of the iteration time of GPT models, followed by formulation for each communication and computation phase.

\begin{table}[t]
	\center
	\small
    \begin{tabular}{|p{1.2cm}|p{6.5cm}|}
        \hline
        \textbf{Notation} & \textbf{Explanation} \\
        \hline
        $p, t, d$ & $(p,t,d)$ for the pipeline-, tensor-, and data-parallel degree, respectively. \\
        \hline
        $N$ & Total number of model parameters. \\
        \hline
        $l$ & Number of transformer block layers\\
        \hline
        $h$ & Hidden size  \\
        \hline
        $s$ & Sequence length \\
        \hline 
        $g, b$ & Global batch size and micro-batch size \\
        \hline
        $m$ & Number of micro-batches \\
        \hline
        $C_{TP}$, $C_{PP}$, $C_{DP}$ & Effective bandwidth during TP, PP, and DP communication phases, respectively. \\
        \hline
        $F$ & GPU's maximum computation capacity (\ie, peak FP16 FLOP/s) \\
        \hline
        $\mu$ & GPU's effective utilization rate during computation \\
        \hline
        \hline
        $e$ & Expert parallel degree\\
        \hline
        $C_{ATA}$ & Effective bandwidth during AllToAll communication phase. \\
        \hline
    \end{tabular}
    \caption{Notations for analysing GPT and MoE models. The last two are specific notations for MoE models.}
    \label{table:notation}
\end{table}

\begin{figure}[t]
	\centering
	\includegraphics[width=\linewidth]{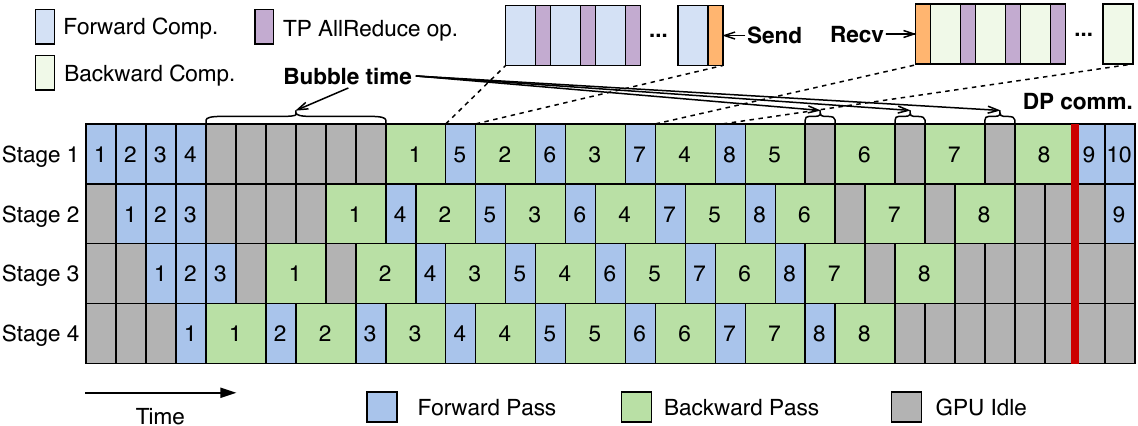}
	\caption{Illustration of iteration time dissection for GPT models under the typical "1F1B" pipeline scheduling with 4 pipeline stages and 8 micro-batches.}
	\label{fig:pipeline}
\end{figure}

\begin{table*}[t]
	\center
	\small 
	\begin{tabular}{|c|c|c|c|c|c|c|c|c|c|c|}
	\hline
	\textbf{\thead{No. of \\ parameters (B)}} & \textbf{\thead{No. of \\ layers}} & \textbf{\thead{Attention \\ heads}} & \textbf{\thead{Hidden \\ size}} & \textbf{\thead{Sequence \\ length}} &  \textbf{\thead{(t, p, d) \\ parallel size}} & \textbf{\thead{Batch \\ size}} & \textbf{\thead{Max. global \\ batch size}} & \textbf{\thead{GPU series (Mem size; FLOP/s)}} \\
	\hline
	1.5 & 48 & 16 & 1600 & 1024 & (2, 4, 2) & 256 & 512 & GeForce (24GB; 71T) \\
	\hline
	1.5 & 48 & 16 & 1600 & 1024 & (0, 2, 2) & 256 & 512 & Tesla (32GB; 112T) \\
	\hline
	3 & 54 & 20 & 2000 & 1024 & (2, 4, 2) & 256 & 512 & GeForce (24GB; 71T) \\
	\hline
	3 & 54 & 20 & 2000 & 1024 & (2, 2, 2) & 256 & 512 & Tesla (32GB; 112T) \\
	\hline
	39 & 48 & 64 & 8192 & 2048 & (4, 4, 2) & 48 & 1536 & Hopper (80GB; 989T) \\
	\hline
	39 & 48 & 64 & 8192 & 2048 & (8, 2, 2) & 48 & 1536 & Hopper (80GB; 989T) \\
	\hline
	76 & 60 & 80 & 10240 & 2048 & (4, 8, 2) & 64 & 1793 & Hopper (80GB; 989T) \\
	\hline
	145 & 80 & 96 & 12288 & 2048 & (8, 8, 1) & 96 & 2304 & Hopper (80GB; 989T) \\
	\hline
	\end{tabular}
	\caption{Model configurations for GPT models from 1.5B to 145B parameters.}
	\label{tab:model-performance}
\end{table*}

\parab{Iteration time.} As depicted in Fig.~\ref{fig:pipeline}, the TP \texttt{AllReduce} operation is executed multiple times during the processing of each micro-batch. PP \texttt{Send} and \texttt{Recv} operations occur at the boundaries of pipeline stages, while the DP \texttt{AllReduce} operation takes place at the end of each iteration. The total iteration time can be defined as the complete working time of stage $1$, which completes the backward computation of the last micro-batch at the latest. Assuming that a GPU indexed as $rank_0$ is at stage 1, the iteration time can be further defined as the sum of computation, communication (including TP, PP, and DP), and bubble time at $rank_0$:
\begin{equation}
T_{iter} = T_{comp} + T_{TP} + T_{PP} + T_{DP} + T_{bubble}
\label{eq:iter}
\end{equation}

Note that the DP \texttt{AllReduce} could potentially overlap with the backward computation. However, DP is insignificant in the training of large GPT models due to its low traffic volume and minimal contribution to the iteration time, as shown in Fig.~\ref{fig:volume}. Hence, our analysis assumes no overlap between DP communication and backward computation for simplicity. 

\parab{TP time.} Each TP \texttt{AllReduce} operation induces $2bsh$ bytes of traffic within a TP group. If Ring-AllReduce is adopted, each \texttt{AllReduce} actually generates $2bsh \times \frac{2(t-1)}{t}$ bytes of traffic. There are four such operations per micro-batch and transformer block. Given that the recomputation (by default enabled) adds two additional \texttt{AllReduce}, the total comes to six \texttt{AllReduce} operations. With the number of transformer blocks allocated to one stage being $l/p$, the TP time per iteration is calculated as follows, where $T_{TP}^{mb}$ represents the TP time for one micro-batch:
\begin{equation}
T_{TP} = m \times T_{TP}^{mb} = m \times \frac{l}{p} \times  \frac{6 \times 2bsh \times 2(t-1)}{t \times C_{TP}}
\end{equation}

\parab{PP time.} Each PP \texttt{Send}/\texttt{Recv} operation transfers $2bsh$ bytes of data between two GPUs in adjacent pipeline stages. At $rank_0$, one \texttt{Send} and one \texttt{Recv} operation occur per micro-batch, leading to the following formulation for PP time:
\begin{equation}
T_{PP} = m \times T_{PP}^{mb} = m \times \frac{2 \times 2bsh}{C_{PP}}
\end{equation}

\parab{DP time.} After completing all backward computations, a DP \texttt{AllReduce} is used to aggregate model parameters. Assuming a uniform distribution of model parameters ($N$) across pipeline stages (a simplification, as the actual distribution may slightly vary), the DP time is calculated as:
\begin{equation}
T_{DP} = \frac{2N}{p \times t} \times \frac{2(d-1)}{d \times C_{DP}}
\end{equation}

\parab{Bubble time.} Considering the computation time for one micro-batch, including both forward and backward passes, as $T_{comp}^{mb}$, we can derive the formulation of bubble time under the "1F1B" pipeline scheduling scheme as:
\begin{equation}
T_{bubble} = (p-1) \times (T_{comp}^{mb}+T_{PP}^{mb} +T_{TP}^{mb})
\end{equation}

Furthermore, in addition to Eq.~\ref{eq:iter}, the total iteration time can also be given by:
\begin{equation}
T_{iter} = m \times (T_{comp}^{mb}+T_{PP}^{mb} +T_{TP}^{mb}) + T_{bubble} + T_{DP}
\end{equation}

Since $T_{DP}$ is negligible during GPT model training, we can exclude it. Therefore, the ratio of bubble time, $R_{bubble}$, can be approximated as a constant value:
\begin{equation}
R_{bubble} = T_{bubble}/T_{iter} \approx (p-1)/(p-1+m)
\label{eq:bubble-ratio}
\end{equation}


\parab{Computation time.} It is estimated that each model parameter and input token requires roughly eight floating-point operations (FLOPs) for computation, two for the forward pass, two for the recomputation, and four for the backward pass~\cite{narayanan2021efficient,korthikanti2023reducing}. Adopting this value, the computation requirement for each micro-batch is calculated as: 
\begin{equation}
FLOPs^{mb} = 8 \times \frac{N}{p\times t} \times b \times s
\end{equation}

The computation time is then the ratio of required FLOPs to the GPU's capacity ($F$). Given the unattainability of the GPU's peak computational capacity, in practice, we introduce a factor $\mu$ to represent the GPU utilization rate. Thus, the computation time per iteration is estimated by: 
\begin{equation}
T_{comp} = \frac{m \times FLOPs^{mb}}{\mu F} = \frac{8m \times N \times b \times s}{p \times t \times \mu F}
\end{equation}

\begin{table}[t]
	\center
	\small
    \begin{tabular}{|p{1.6cm}|p{2.7cm}|p{2.7cm}|}
        \hline
        \textbf{GPU series} & \textbf{Intra-node network} & \textbf{Inter-node network} \\
        \hline
        GeForce & PCIe3.0x16 (32GB/s) & one 100Gb CX5\\
        \hline
        Tesla (PCIe) & PCIe3.0x16 (32GB/s) & one 100Gb CX5\\
        \hline
        Tesla (NVL) & NVLink (300GB/s) & two 100Gb CX5 \\
        \hline
        Hopper & NVSwitch (400GB/s) & eight 400Gb CX7 \\
        \hline
    \end{tabular}
    \caption{Specifications of four hardware platforms. }
    \label{table:platform}
\end{table}

\begin{figure*}[t]
	\centering
	\begin{subfigure}[t]{0.42\linewidth}
		\includegraphics[width=\linewidth]{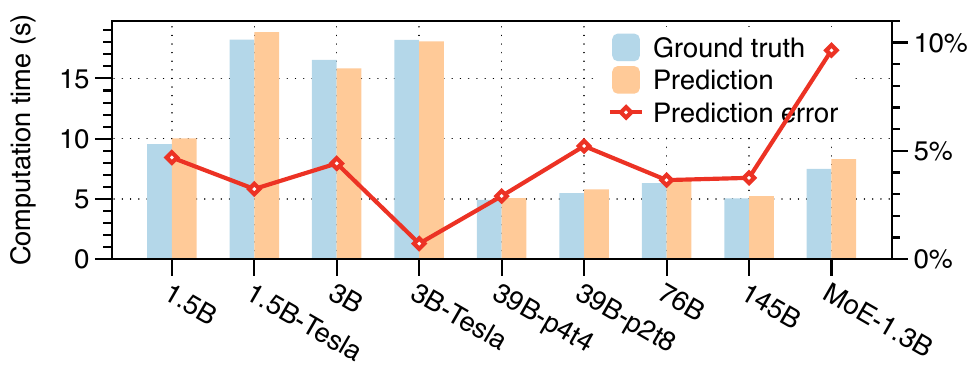}
		\caption{Computation time. The unit of $y$ axis is second.}
		\label{fig:formula:1}
	\end{subfigure}
	\begin{subfigure}[t]{0.42\linewidth}
		\includegraphics[width=\linewidth]{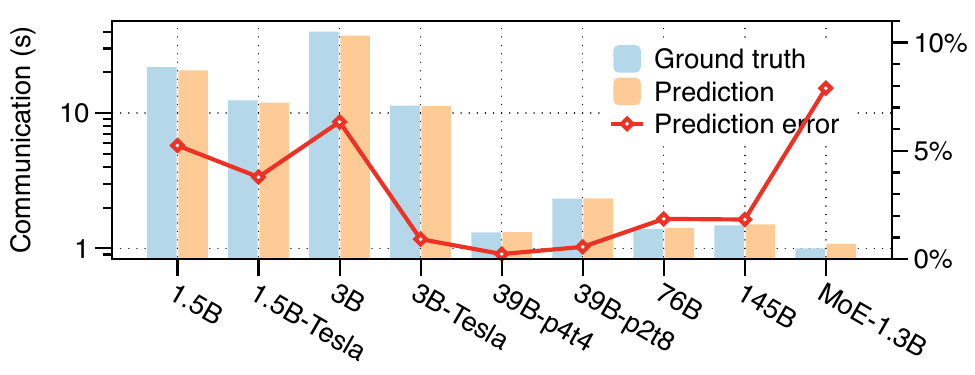}
		\caption{Communication time. The unit of $y$ axis is second.}
		\label{fig:formula:2}
	\end{subfigure}
	\begin{subfigure}[t]{0.42\linewidth}
		\includegraphics[width=\linewidth]{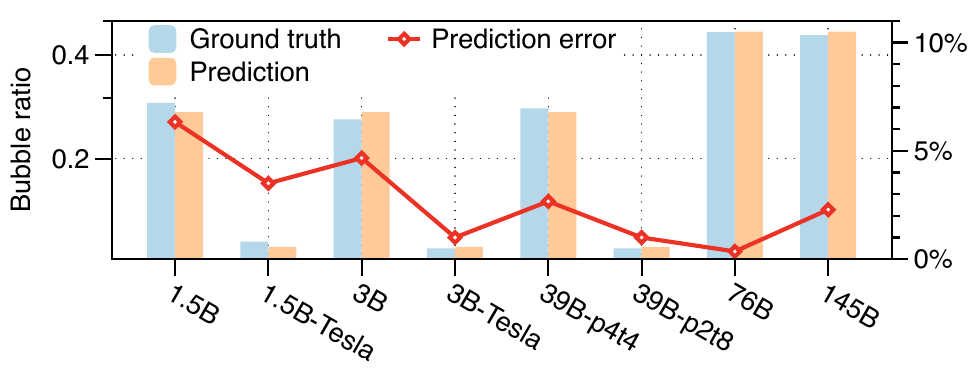}
		\caption{Bubble ratio.}
		\label{fig:formula:3}
	\end{subfigure}
	\begin{subfigure}[t]{0.42\linewidth}
		\includegraphics[width=\linewidth]{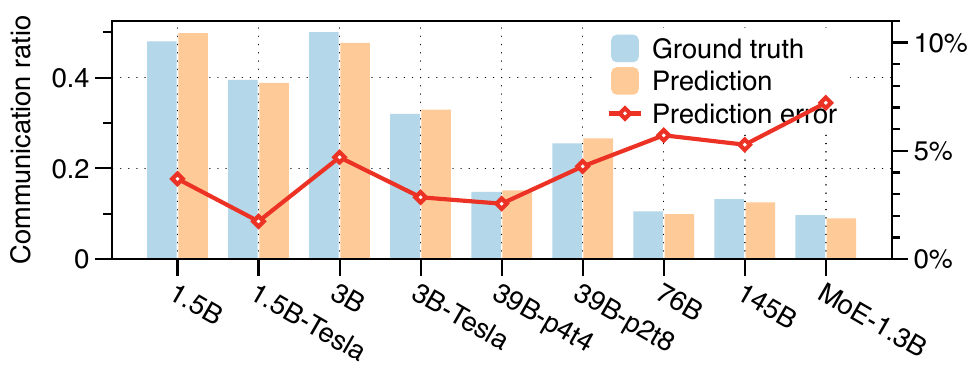}
		\caption{Communication ratio.}
		\label{fig:formula:4}
	\end{subfigure}
	\caption{Comparision between our formulation's predictions and empirical results (ground truth) for eight GPT models with various model sizes and system settings. Our formulation shows satisfactory accuracy.}
	\label{fig:formula}
\end{figure*}

\parab{Impact of interleaved 1F1B schedule.} The interleaved 1F1B schedule~\cite{narayanan2021efficient} enables each GPU to perform computation for multiple subsets of layers (called a model chunk), instead of a stage (a single contiguous set of layers). Assuming a stage is split into $v$ model chunks, the interleaved 1F1B can reduce the bubble ratio by $v$ times: $R_{bubble}^{interleaved} = R_{bubble}/v$, and increase the PP time by $v$ times: $T_{PP}^{interleaved} = v \times T_{PP}$, compared to the original 1F1B schedule.

\subsection{Approximate Modeling for AllToAll}\label{formulation:moe}
Compared to GPT models, the significant distinction within MoE models lies in the AllToAll communication for expert layers. Aside from this, other communication and computation phases for non-expert layers in MoE models are similar to those in GPT models. Consequently, we primarily focus on the formulation of AllToAll communication phases.

In MoE models, an MLP layer is replaced into an expert layer for every two transformer blocks (a common configuration~\cite{ds}). There are \textit{six} AllToAll operations for each expert layer: two operations each for the forward, the backward, and the recomputation pass, respectively. The total byte size of the token volume is $2gsh$ with a maximum traffic volume of $2 \times 2gsh$ under an adaptive top-2 gating mechanism, as each token selects a maximum of two experts. Assuming a uniform selection of experts with an expert parallel degree of $e$, the traffic volume on each GPU pair during a single AllToAll operation is $\frac{2 \times 2gsh}{e}$. Therefore, the total AllToAll time during one iteration is calculated as:
\begin{equation}
	T_{ATA} = \frac{l}{2} \times \frac{6 \times 2 \times 2gsh}{e \times C_{ATA}}
\label{eq:alltoall}
\end{equation}

\subsection{Evaluation of Estimation Accuracy}\label{formulation:vali}
\parab{GTP models.} We conduct realistic training of eight GPT models ranging from 1.5B to 145B parameters. The specifications of these models are provided in Table~\ref{tab:model-performance}. The training was performed on three types of GPUs from the GeForce, Tesla, and Hopper series, with hardware specifications shown in Table~\ref{table:platform}. Note that the 1.5B and 3B models are evaluated on both GeForce and Tesla GPUs, while the 39B model is evaluated using two parallelism configurations. The configurations of the per-replica batch size for 1.5B and 3B models follow instructions in \cite{radford2019language} while other models follow instructions in \cite{brown2020language}. We employ the Megatron-DeepSpeed framework~\cite{ds} to deploy the hybrid parallel training for GPT models, with NCCL~\cite{nccl} as the communication backend. 

To assess the accuracy of the analytical formulation, we compare its predictions with empirical data from our experiments (ground truth). Our formulation breaks down the iteration into several phases, including computation time, communication time (covering TP, PP, and DP), and bubble ratio. We first evaluate the estimates for each phase and then assess the communication ratio ($R_{comm} = T_{comm}/T_{iter}$), which is commonly used to represent communication overhead. 

In our formulation, all variables are derived directly from model and hardware specifications, except for the GPU utilization rate during the computation phase ($\mu$) and effective bandwidth during the communication phase ($C_{TP}$, $C_{PP}$, $C_{DP}$). For $\mu$ and $C$, we integrate benchmarks from the GeForce, Tesla, and Hopper series GPUs as reference values. Given the stability of $\mu$ and $C$, these benchmarks provide reliable reference points that can be effectively incorporated into our formulation. Further details about these variables are in $\S$\ref{benchmark}. Note that we adopt the form in Eq.~\ref{eq:bubble-ratio} for estimating the $R_{bubble}$.

As depicted in Fig.~\ref{fig:formula}, our analytical formulation demonstrates approximately 95\% accuracy in estimating the computation time, communication time, bubble ratio, and communication ratio for GPT models, compared to the ground truth. This level of accuracy is maintained across the majority of our experiments. Fig.~\ref{fig:formula}\violet{b} \& Fig.~\ref{fig:formula}\violet{d} illustrate the time and ratio for communication phase, with the prediction error indicating the prediction gap. Notably, TP time covers 85\%$\sim$99\% of the total communication time and 10\%$\sim$45\% of the iteration time. In contrast, the communication times for DP and PP are minimal, accounting for only about 1\% of the iteration time, reflective of their relatively low traffic volumes.


\parab{MoE models.} Using the estimation formula from Eq.~\ref{eq:alltoall}, we predict the total AllToAll communication volume for the MoE-1.3B model. Fig.~\ref{fig:formula}\violet{a} to Fig.~\ref{fig:formula}\violet{d} illustrates the computation time, communication time, and ratio for this MoE-1.3B model. As results show, the prediction error is around 8\% to 10\%, slightly higher than the GPT models.

\parab{Conclusions applicable to larger scales.} Due to the limited number of available GPUs, most of our experiments were conducted with a data-parallel degree of two. However, the core conclusions remain applicable to larger-scale training with more data-parallel groups. This broader applicability is due to the consistently minimal contribution of DP time to the overall iteration time at larger scales. The minimal impact of DP communication is attributed to two factors: the traffic volume during DP communication (which is constrained by GPU memory capacity) and the effective bandwidth during DP communication, both of which remain constant across varying scales. We further evaluate the inter-node AllReduce communication with varying scales in $\S$\ref{bench:inter-node}.

 
\parab{Takeaway \#3:} \textit{This analytical formulation precisely models the whole training process of GPT models, offering a mathematic understanding of predictability. The AllToAll communication phase of MoE training can be approximately predicted due to its semi-predictability.}

\section{Analyzing Critical Factors in Formulation}\label{benchmark}
The GPU utilization rate ($\mu$) during the computation phase and the effective bandwidth ($C$) during communication operations are critical factors of the formulation in $\S$\ref{modeling}. This section presents comprehensive benchmarks of them, where we vary the determinants and assess the influence. 

\subsection{Experimental setup}
We evaluate four hardware platforms, referred to as GeForce, Tesla-PCIe, Tesla-NVLink, and Hopper, each with distinct features such as GPU series, computational capacity, memory size, and interconnect topology, as detailed in Table~\ref{table:platform}.  Within each platform, we range various determinants to evaluate their impact on $C$ and $\mu$. 

\parab{Intra-node GPU interconnect.} The GeForce and Tesla-PCIe platforms use \textit{PCIe}, while Tesla-NVLink and Hopper platforms utilize \textit{NVLink} and \textit{NVSwitch} interconnect topologies, respectively. These three types of GPU interconnect topologies are detailedly described in Appendix~\ref{apx:topo}.

\parab{Inter-node GPU interconnect.} The inter-node network topology for the GeForce, Tesla-PCIe, and Tesla-NVLink platforms utilizes a Top of Rack (ToR) architecture, in which all machines connect to a single ToR switch. In contrast, the Hopper platform employs the standard NVIDIA DGX SuperPOD infrastructure~\cite{superpod}, which is a typical Leaf-Spine topology.


\begin{figure}[t]
	\centering
	\begin{subfigure}[t]{0.44\linewidth}
		\includegraphics[width=\linewidth]{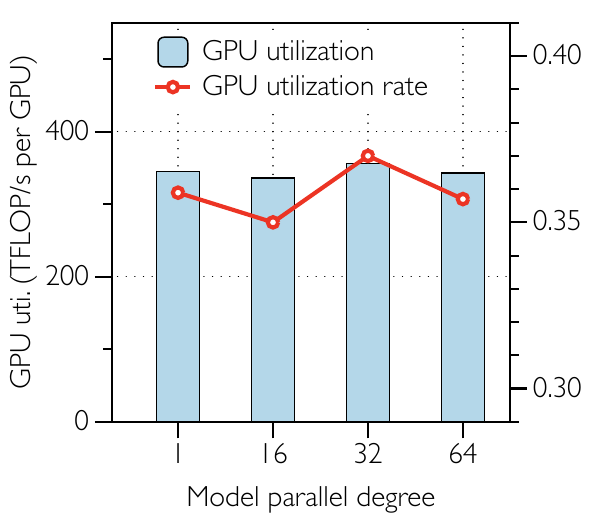}
		\caption{Stable rate.}
		\label{fig:comp-consis}
	\end{subfigure}
	\begin{subfigure}[t]{0.55\linewidth}
		\includegraphics[width=\linewidth]{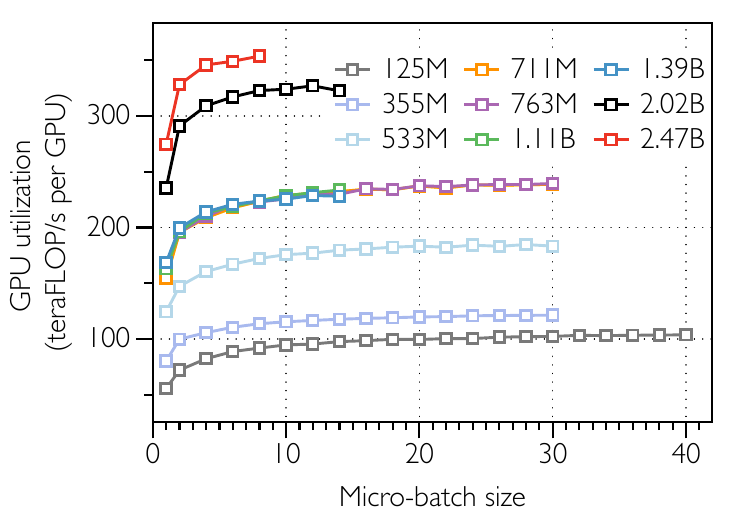}
		\caption{Empirical values.}
		\label{fig:comp-empirical}
	\end{subfigure}
	\caption{GPU utilizations during the computation phase remains stable across varying parallel scales (a) but varies with varying per-GPU model sizes and micro-batch sizes (b).}
	\label{fig:comp}
\end{figure}

\subsection{Analysis on GPU Utilization}\label{bench:uti}
We first demonstrate that, with a fixed per-GPU model size and micro-batch size, the GPU utilizations remains consistent across various distributed scales. Then, we benchmark the GPU utilization under different per-GPU model sizes and micro-batch sizes. All four platforms show consistent results; thus we only present the results from the Hopper platform.

\parab{Consistent GPU utilization.} It is observed that the GPU utilization during the micro-batch computation phase remains stable across different scales of distributed training when the per-GPU model size and micro-batch size are fixed. To illustrate, we measured the average GPU utilization and rate under four training scales: one involving a single GPU and the other employing distributed training with model parallel sizes of 16, 32, and 64. In each scenario, the per-GPU model size is set at approximately 2.4 billion parameters, and the micro-batch size is maintained at 6. The results, depicted in Fig.~\ref{fig:comp-consis}, confirm that the GPU utilization and rate remain relatively consistent across the four training scales. This consistency allows that, as long as we maintain the same per-GPU model size and micro-batch size, the measured values from the single GPU training scenario can serve as reliable reference points for larger distributed training configurations.

\parab{Empirical values of GPU utilization.} We next explore the impact of varying per-GPU model sizes and micro-batch sizes on GPU utilization under the single GPU training scenario. Adjusting these parameters, we measure the corresponding GPU utilization, shown in Fig.~\ref{fig:comp-empirical}. These findings confirm that both per-GPU model size and micro-batch size significantly influence GPU utilization. These measured empirical values can serve as reference points of the GPU utilization rate for the formulation in $\S$\ref{modeling}. By incorporating empirical values that align with the specific per-GPU model size and micro-batch size into the formulation, we can achieve precise predictions in distributed training scenarios.

\parab{Takeaway \#4:} \textit{The GPU utilization remains consistent given fixed per-GPU model size and micro-batch size. Thus, the empirically measured GPU utilizations at the single GPU training scenario can serve as reference points for formulations in distributed training scenarios.}

\begin{figure}[t]
	\centering
	\begin{subfigure}[t]{0.325\linewidth}
		\includegraphics[width=\linewidth]{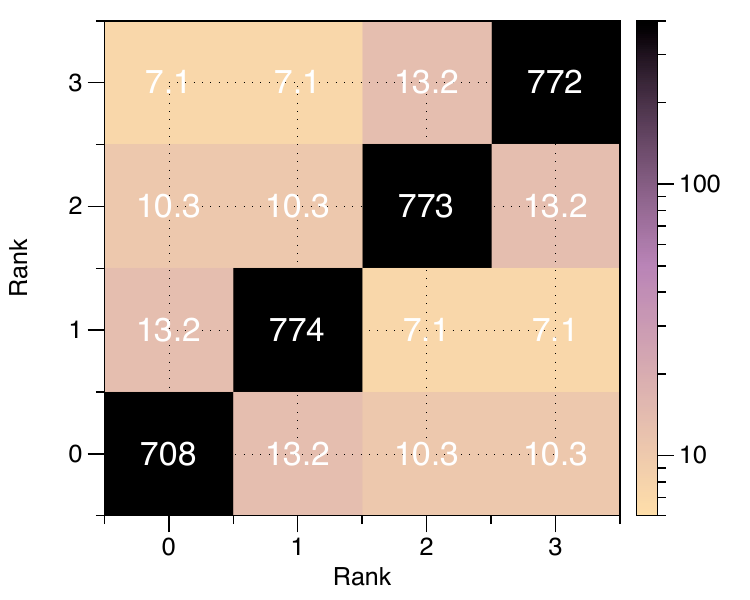}
		\caption{PCIe bandwidth.}
	\end{subfigure}
	\begin{subfigure}[t]{0.325\linewidth}
		\includegraphics[width=\linewidth]{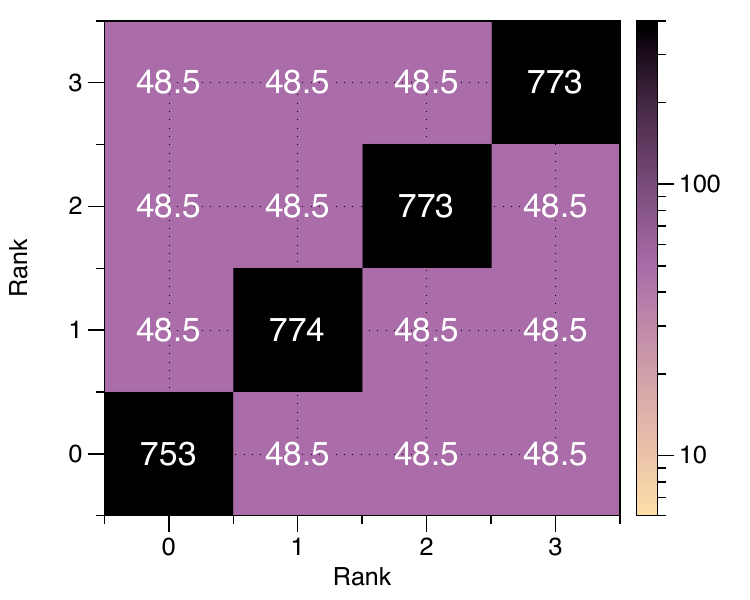}
		\caption{NVLink BW.}
	\end{subfigure}
	\begin{subfigure}[t]{0.325\linewidth}
		\includegraphics[width=\linewidth]{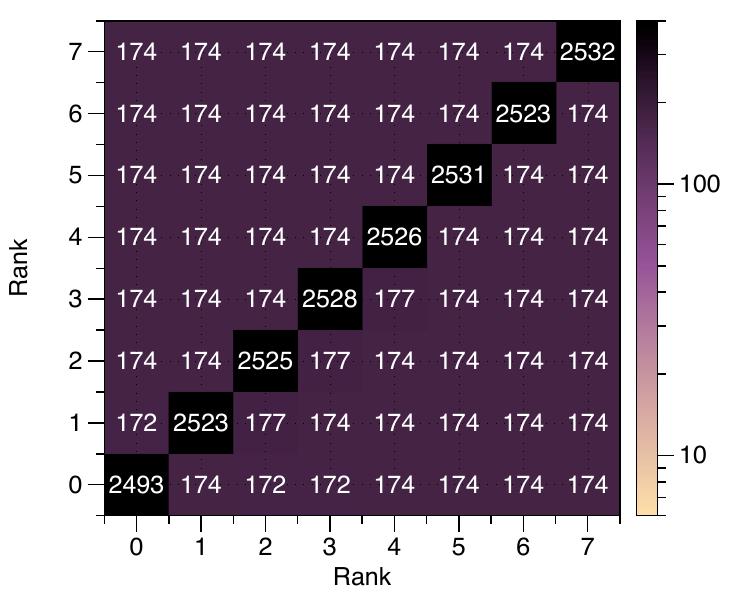}
		\caption{NVSwitch BW.}
	\end{subfigure}
	\begin{subfigure}[t]{0.325\linewidth}
		\includegraphics[width=\linewidth]{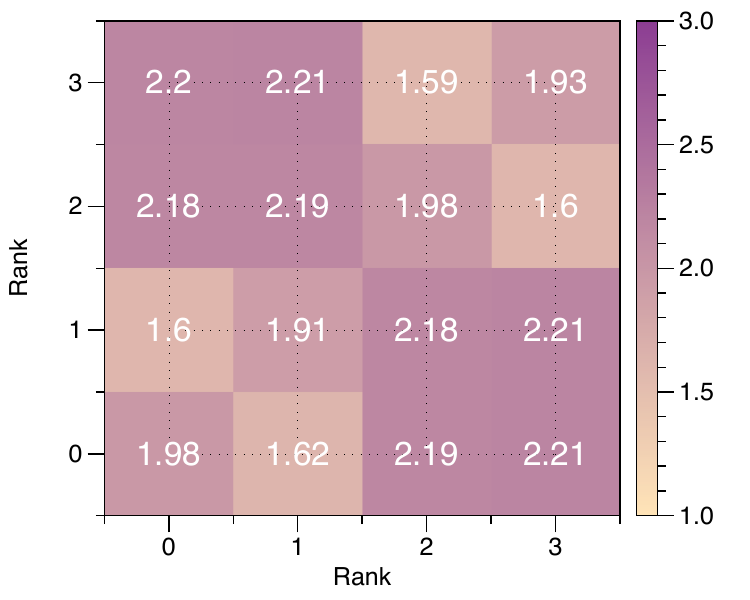}
		\caption{PCIe latency.}
	\end{subfigure}
	\begin{subfigure}[t]{0.325\linewidth}
		\includegraphics[width=\linewidth]{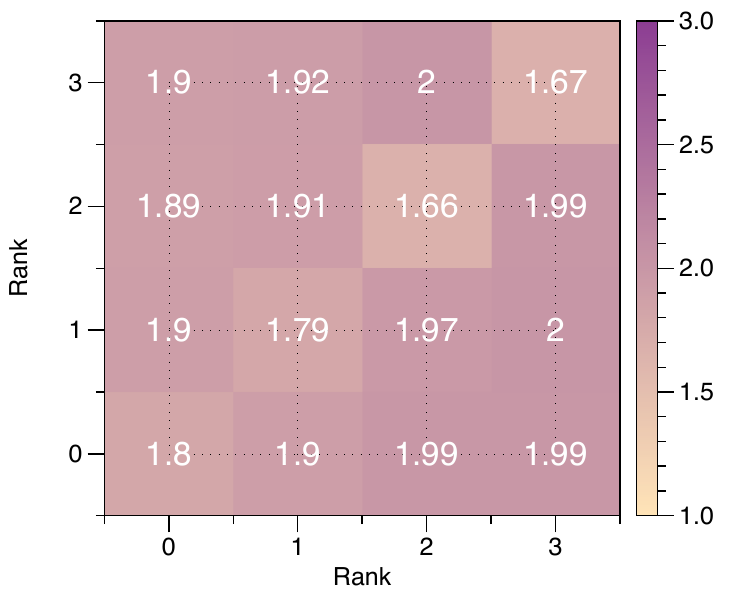}
		\caption{NVLink latency.}
	\end{subfigure}
	\begin{subfigure}[t]{0.325\linewidth}
		\includegraphics[width=\linewidth]{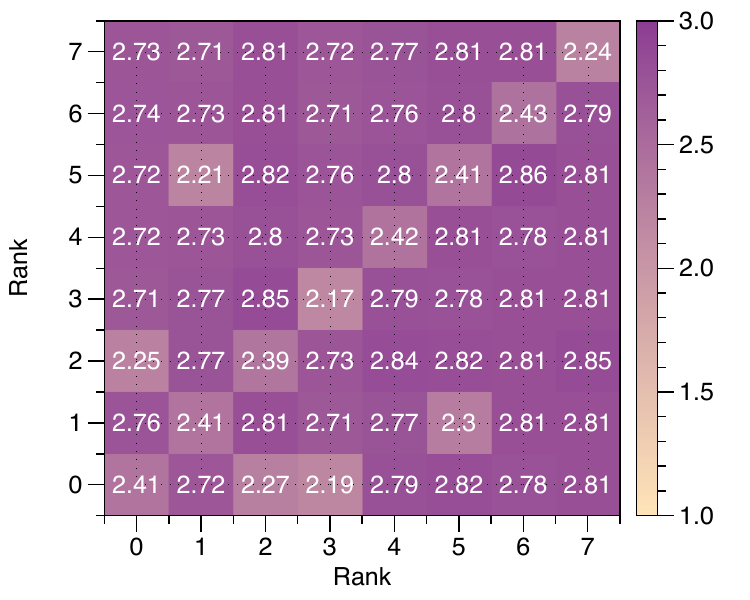}
		\caption{NVSwitch latency.}
	\end{subfigure}
	\caption{Unidirectional bandwidth and start-up latency for intra-node communications under three topologies.}
	\label{fig:p2pintra}
\end{figure}

\begin{figure*}[t]
	\centering
	\begin{subfigure}[t]{0.3\linewidth}
		\includegraphics[width=\linewidth]{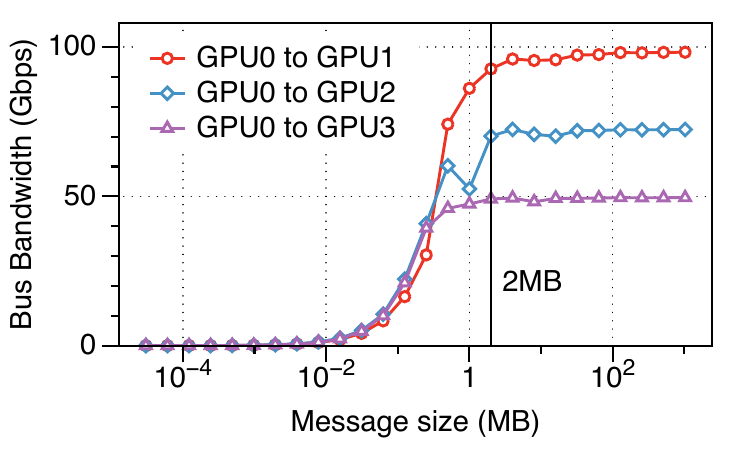}
		\caption{PCIe.}
	\end{subfigure}
	\begin{subfigure}[t]{0.3\linewidth}
		\includegraphics[width=\linewidth]{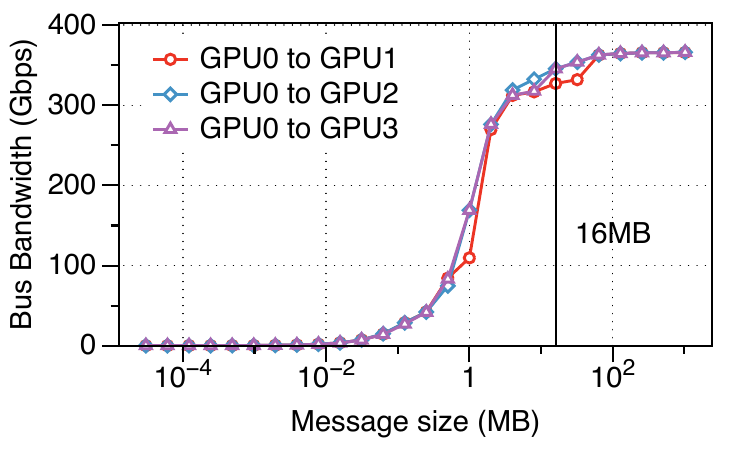}
		\caption{NVLink.}
	\end{subfigure}
	\begin{subfigure}[t]{0.3\linewidth}
		\includegraphics[width=\linewidth]{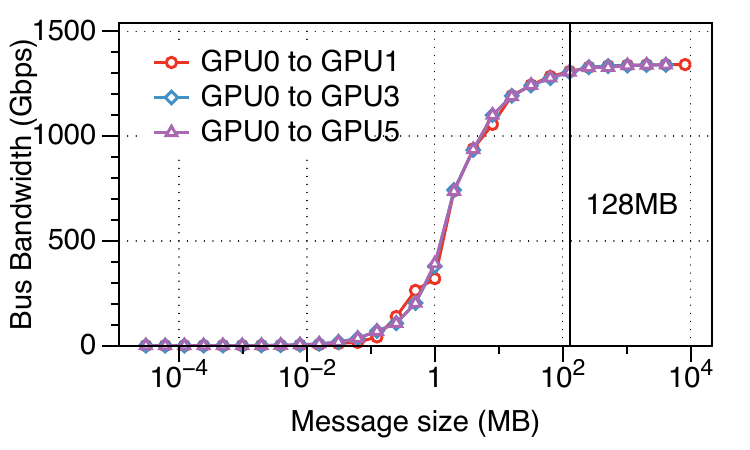}
		\caption{NVSwitch.}
	\end{subfigure}
	\caption{Effective bandwidth of intra-node P2P communication with varying message size under the PCIe, NVLink and NVSwitch GPU interconnect topologies.}
	\label{fig:intra}
\end{figure*}

\begin{figure*}[t]
	\centering
	\begin{subfigure}[t]{0.3\linewidth}
		\includegraphics[width=\linewidth]{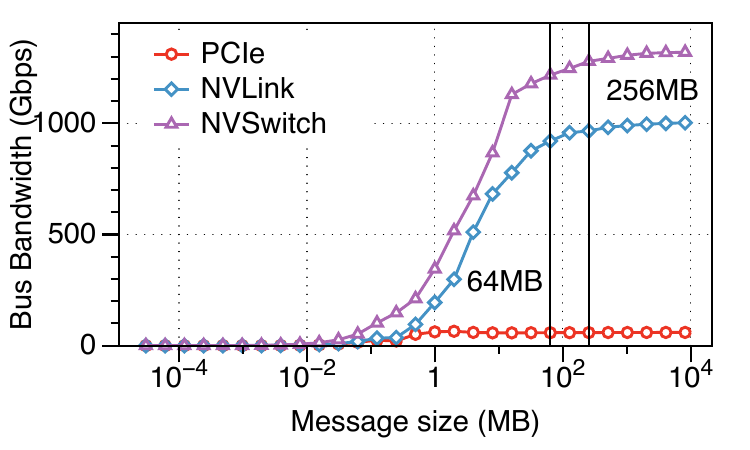}
		\caption{Three topologies (fixed scale).}
		\label{fig:intra-allreduce}
	\end{subfigure}
	\begin{subfigure}[t]{0.3\linewidth}
		\includegraphics[width=\linewidth]{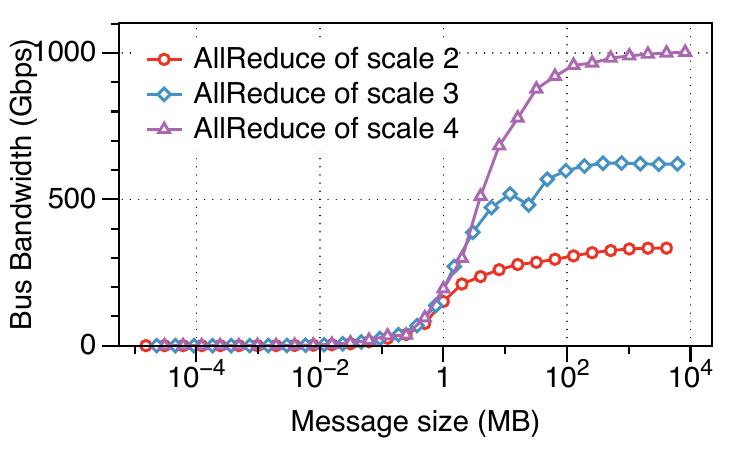}
		\caption{NVLink with varying scales.}
		\label{fig:mutipath:1}
	\end{subfigure}
	\begin{subfigure}[t]{0.3\linewidth}
		\includegraphics[width=\linewidth]{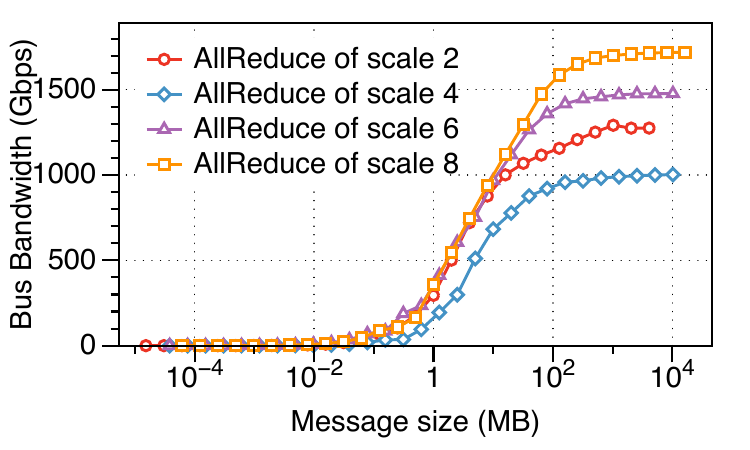}
		\caption{NVSwitch with varying scales.}
		\label{fig:mutipath:2}
	\end{subfigure}
	\caption{Effective bandwidth of intra-node AllReduce communication under three topologies (a); Intra-node AllReduce communication exhibit obvious NUMA effects under NVLink (b) and NVSwitch (c) topologies.}
	\label{fig:mutipath}
\end{figure*}

\subsection{Analysis on Effective Bandwidth}\label{bench:inter-node}
We comprehensively analyze the impact of various factors on the effective bandwidth of P2P and collective operations. These factors include GPU interconnect topology, message size, collective scale, synchronization overhead, \etc.

\parab{GPU interconnect topology sets the maximum potential.} Fig.~\ref{fig:p2pintra} illustrates the sustainable unidirectional bandwidth and startup latency for intra-node communications across three different topologies: PCIe, NVLink, and NVSwitch. Notably, the colorbar units for bandwidth and latency are GB/s and $\mu$s, respectively. In terms of unidirectional bandwidth, the highest P2P bandwidth for PCIe, NVLink, and NVSwitch are 13.2 GB/s, 48.4 GB/s, and 174 GB/s, respectively. These values align with the maximum capacities of PCIe 3.0x16, 2 NVLink lines, and 8 NVLink lines. Furthermore, the PCIe topology shows distinct NUMA effects: GPU pairs within the same PCIe switch achieve higher bandwidth, while those across PCIe switches experience significantly lower bandwidth due to the slower QPI path compared to PCIe paths. In contrast, P2P bandwidth is considerably more uniform in the NVLink and NVSwitch topologies. Regarding the startup latency, all three topologies exhibit slight NUMA effects, though the differences are relatively minor.

\subsubsection{Intra-node Communication}

\parab{P2P communication:} Fig.~\ref{fig:intra} illustrates the effective bandwidth of P2P communication for varying message sizes under the three interconnect topologies. For each topology, several GPU pairs are evaluated. The results indicate that different GPU pairs exhibit different maximum bus bandwidths under the PCIe topology. In contrast, the bandwidth is uniform across all pairs in the NVLink and NVSwitch topologies, consistent with the findings in Fig.~\ref{fig:p2pintra}. Additionally, the results demonstrate that the PCIe, NVLink, and NVSwitch topologies require message sizes of 2MB, 16MB, and 128MB, respectively, to fully utilize the intra-node interconnect capacity. For message sizes below these thresholds, it is the startup latency that predominantly limits the effective bus bandwidth.

\parab{Collective communication with varying message sizes:} Fig.\ref{fig:intra-allreduce}  presents the effective bandwidth of AllReduce communication for varying message sizes across the three interconnect topologies. The collective scale for AllReduce is set to 4 for all topologies. The maximum bus bandwidth of AllReduce in the PCIe topology is approximately 50Gbps, consistent with the bottleneck capacity of P2P bandwidth (7.1GB/s) in the PCIe topology. The maximum bus bandwidth of AllReduce in the NVLink and NVSwitch topologies are significantly higher than in PCIe, reaching about 1000Gbps and 1500Gbps, respectively, which are slightly lower than the total capacities of 6 NVLink slots (1200Gbps) and 8 NVLink slots (1600Gbps) in the NVLink and NVSwitch topologies. This bandwidth gap results from the \textit{synchronization overhead} inherent in collective communications~\cite{ma2022autobyte,peng2019generic}. 

\parab{Collective communication with NUMA effects triggered by collective scales:} The effective bandwidth of AllReduce in NVLink and NVSwitch topologies is impacted by NUMA effects triggered by collective scales; This phenomenon arises from the presence of \textit{multiple paths}. As illustrated in Fig.~\ref{fig:mutipath:1}, AllReduce of scale 4 achieves approximately 3$\times$ and 2$\times$ higher bandwidth than AllReduce of scale 3 and 2, respectively. This increase is due to the utilization of more rings in larger AllReduce groups within the NVLink topology. Conversely, in NVSwitch topologies, AllReduce operations of any size should achieve similar bandwidth. The observed NUMA effects in NVSwitch may be due to NCCL's inefficient adaptation to leverage the multiple available paths effectively (Fig.~\ref{fig:mutipath:2}).

\subsubsection{Inter-node Communication}
We vary message sizes and collective scales and investigate their impact on the effective bandwidth for inter-node P2P and AllReduce communications. Notably, the PCIe and NVLink topologies share similar GPU-to-RNIC interconnects, resulting in similar inter-node communication performance. Thus, we present only the results of the NVLink topology.

\begin{figure}[t]
	\centering
	\begin{subfigure}[t]{0.495\linewidth}
		\includegraphics[width=\linewidth]{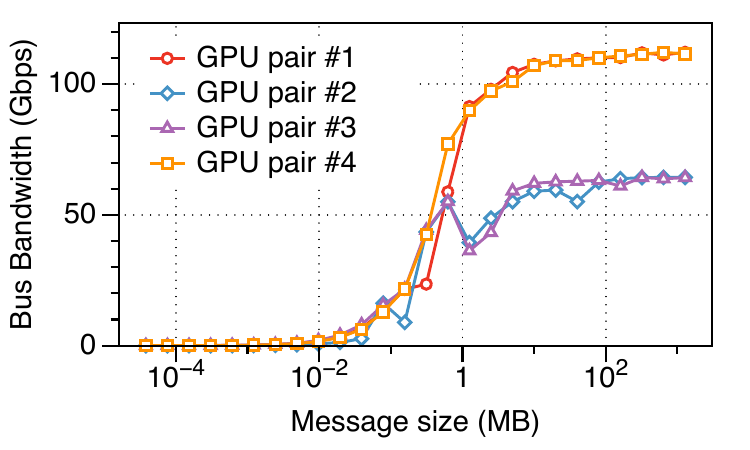}
		\caption{NVLink.}
		\label{fig:inter-p2p-1}
	\end{subfigure}
	\begin{subfigure}[t]{0.495\linewidth}
		\includegraphics[width=\linewidth]{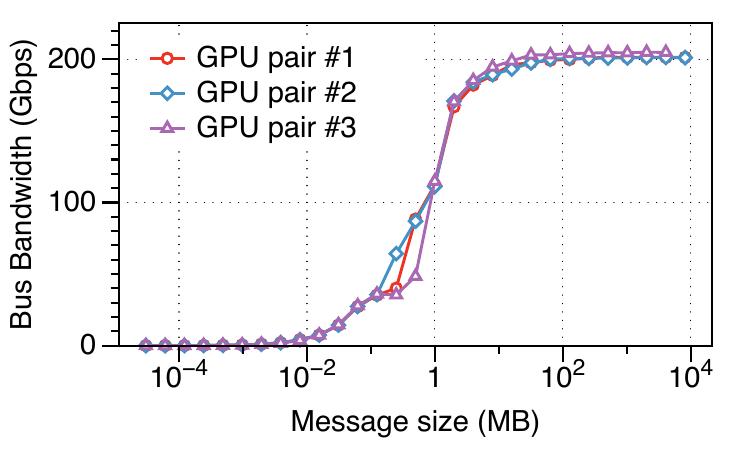}
		\caption{NVSwitch.}
		\label{fig:inter-p2p-2}
	\end{subfigure}
	\caption{Effective bandwidth of inter-node P2P communication in NVLink and NVSwitch topologies.}
	\label{fig:inter-p2p}
\end{figure}

\begin{figure}[t]
	\centering
	\begin{subfigure}[t]{0.495\linewidth}
		\includegraphics[width=\linewidth]{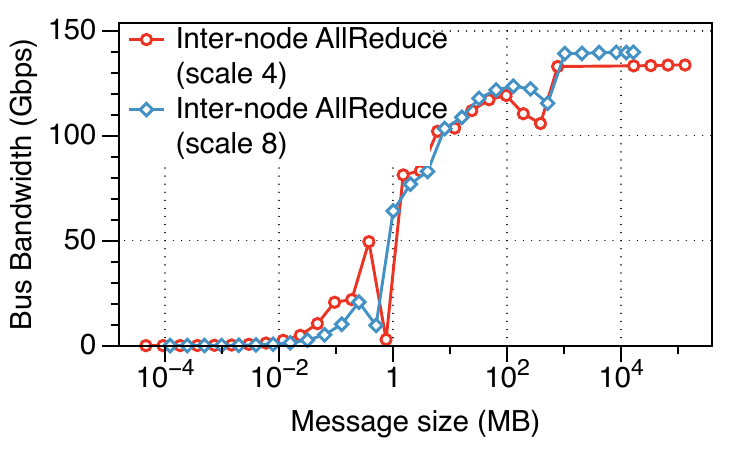}
		\caption{NVLink.}
		\label{fig:inter-ar-1}
	\end{subfigure}
	\begin{subfigure}[t]{0.495\linewidth}
		\includegraphics[width=\linewidth]{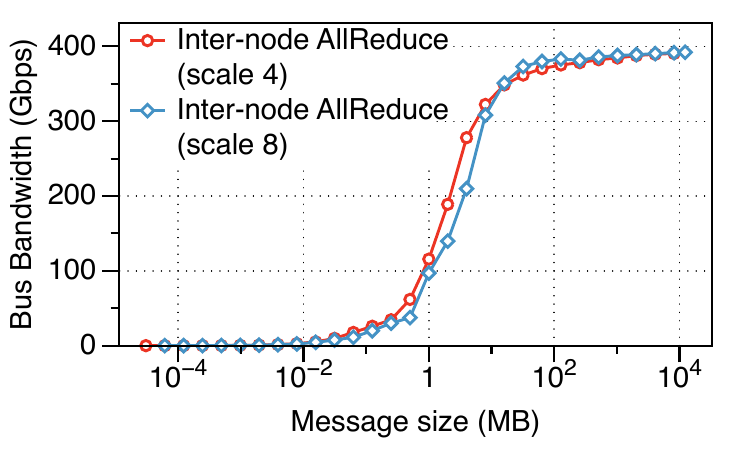}
		\caption{NVSwitch.}
		\label{fig:inter-ar-2}
	\end{subfigure}
	\caption{Effective bandwidth of inter-node AllReduce communication with varying collective scales.}
	\label{fig:inter-ar}
\end{figure}

\parab{P2P Communication.} Fig.~\ref{fig:inter-p2p} illustrates the P2P communication bandwidth across various inter-node GPU pairs in NVLink and NVSwitch topologies. In the NVLink topology, a single RNIC is connected to a PCIe switch. Consequently, the RNIC communicates with GPUs on the same PCIe switch via the PCIe path, and with GPUs under a different PCIe switch via the QPI path. This leads to obvious NUMA effects in its inter-node P2P communication. In contrast, the NVSwitch topology features a directly connected RNIC for every GPU, resulting in uniform inter-node P2P communication across different inter-node GPU pairs.


\parab{Collective communication.} Fig.~\ref{fig:inter-ar} illustrates the effective bandwidth of inter-node AllReduce communication at two different scales (4 and 8) in NVLink and NVSwitch topologies. The participating GPUs are located in different machines, simulating a DP communication process during the training of LLMs. The results indicate that inter-node AllReduce operations, regardless of scale, achieve similar effective bandwidths in both topologies.

\parab{Does congestion exist in single-job training scenario?} During the hybrid training of GPT models, inter-node communications of multiple GPUs occur simultaneously. The Hopper platform assigns one RNIC per GPU, thus eliminating interference among multiple inter-node communications. Although network core congestion among multiple communications is theoretically possible, it was not observed during our experiments. In contrast, as all GPUs share a single RNIC on the GeForce and Tesla platforms, congestion occurs for competing for the RNIC's link capacity. 

\parab{Takeaway \#5:} \textit{The capacity of GPU interconnects sets the maximum potential for the effective bandwidth of communication. Factors such as message size, collective scale, and synchronization overhead are contributors to the lower effective bandwidth than theoretical upper limit.}

\section{Optimizations Guided by Analysis}\label{optimization}
Inspired by the analytical formulation in $\S$\ref{modeling} and the empirical analysis in $\S$\ref{benchmark}, we present a configuration tuning tool called \sys. This tool can predict training performance for various configuration inputs through offline calculations ($\S$\ref{sec:config-tuner}). We showcase the effectiveness of \sys through three optimization scenarios that enhance GPT model training performance ($\S$\ref{sec:tuner-bathsize}-\ref{sec:tuner-dp}).

\begin{figure}[t]
	\centering
	\includegraphics[width=\linewidth]{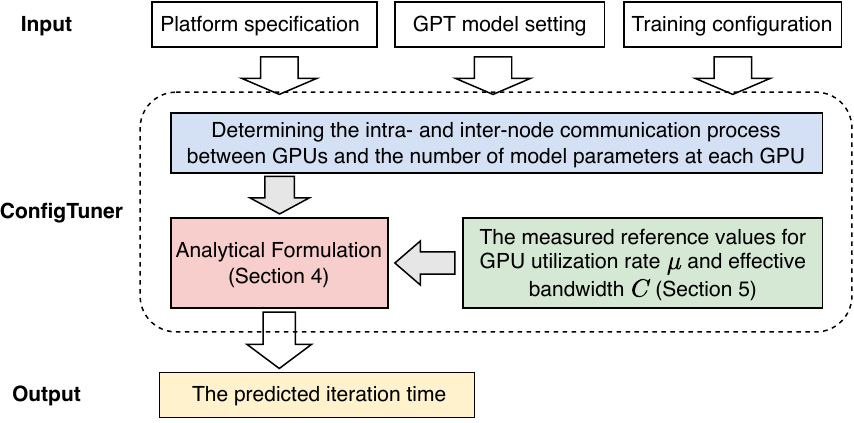}
	\caption{Workflow of \sys.}
	\label{fig:configtuner}
\end{figure}

\subsection{\sys}\label{sec:config-tuner}
An overview of \sys's workflow is illustrated in Fig.~\ref{fig:configtuner}. \sys takes three types of configurations:
\begin{enumerate}
	\item[1.] \textit{Platform Specification}, including the GPU type and intra/inter-node GPU interconnect capacities.
	\item[2.] \textit{GPT Model Setting}, which describes parameters such as the number of layers, hidden size, sequence length, and global batch size. It is important to note that, for a given model size, the global batch size is typically fixed; this is because a smaller global batch size compromises parallel computing performance, while a larger one risks over-fitting~\cite{mccandlish2018empirical}.
	\item[3.] \textit{Training Configuration}, specifying the parallelism degrees (\ie, degrees of tensor, pipeline, and data parallelism) for GPT models, as well as the micro-batch size.
\end{enumerate} 

Once provided with these inputs, \sys calculates the number of parameters each GPU holds and determines the communication matrix and the traffic volumes between all GPUs (with both intra- and inter-node communication). Using the reference values from $\S$\ref{benchmark}, \sys obtains the $C_{TP}$, $C_{PP}$, $C_{DP}$, and $\mu$ variables required by the analytical formulation ($\S$\ref{modeling}). Finally, \sys utilizes this formulation to predict the iteration time (or training throughput such as FLOP/s) as its output. Note that the profiling data required by \sys is minimal, which can be obtained by benchmarking a single GPU and conducting local NCCL tests on a subset of GPUs. In contrast, other configuration tuning tools, such as Alpa~\cite{zheng2022alpa}, typically require running multiple full iterations for each configuration. 

\subsection{Tuning Micro-Batch Size}\label{sec:tuner-bathsize}
We first distinguish between three types of batch sizes used in the training of GPT models: the global batch size, which is the total batch size across all data-parallel model replicas; the batch size, which refers to the batch size of a single model replica (\ie, $\frac{global-batch-size}{data-parallel-degree}$); and the micro-batch size, which is the unit used in pipeline scheduling. Assuming constant GPT model configurations and parallelism degrees, \sys can be used to predict training throughput for different micro-batch sizes, helping us identify the most efficient micro-batch size.

For instance, we evaluate three GPT models with parameter of 39B, 76B, and 145B, using tensor-, pipeline-, and data-parallel degrees of (4, 4, 1), (4, 8, 1), and (8, 8, 1), respectively. The batch sizes per model replica are set to 48, 59, and 96. We limit the data-parallel degree to one because micro-batch size primarily affects the performance of model-parallelism. We use the Hopper platform as the hardware specification and vary the micro-batch size as input to \sys. 

\begin{figure}[t]
	\centering
	\begin{subfigure}[t]{0.49\linewidth}
		\includegraphics[width=\linewidth]{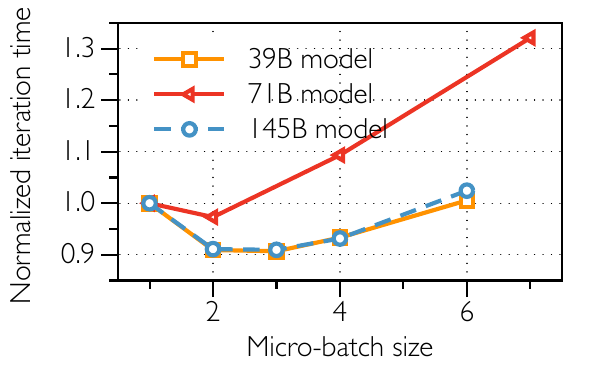}
		\caption{Predictions of \sys.}
		\label{fig:opti-batchsize-1}
	\end{subfigure}
	\begin{subfigure}[t]{0.49\linewidth}
		\includegraphics[width=\linewidth]{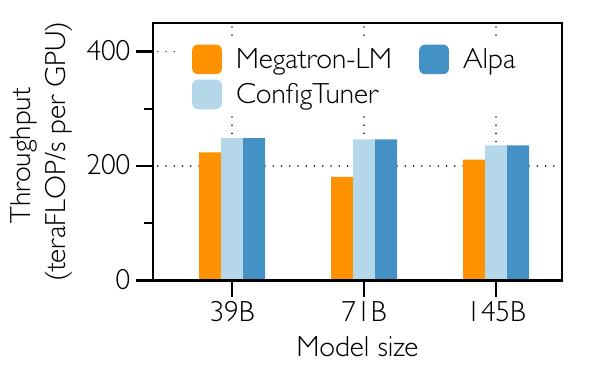}
		\caption{\sys vs. Megatron-LM vs. Alpa}
		\label{fig:opti-batchsize-2}
	\end{subfigure}
	\caption{Tuning micro-batch size using \sys, achieving superior performance than Megatron-LM.}
	\label{fig:opti-batchsize}
\end{figure}

As shown in Fig.~\ref{fig:opti-batchsize-1}, \sys predicts iteration times for different micro-batch sizes. Based on these predictions, we determine the optimal micro-batch sizes to be 3, 2, and 3 for the 39B, 76B, and 145B models, respectively. In contrast, Megatron-LM recommends a micro-batch size of 6, which is the maximum size supported by GPU memory (80GB in our Hopper platform) and is commonly used to maximize GPU utilization~\cite{narayanan2021efficient}. Alpa conducts full training for each configuration and selects the best one. We then conducted real-world experiments, and Fig.~\ref{fig:opti-batchsize-2} shows the empirically measured throughput for the micro-batch sizes chosen by \sys,  Megatron-LM and Alpa. \sys and Apla generate the same configuration suggestion. The results show that \sys's selections achieve throughput improvements of 1.12$\times$, 1.36$\times$, and 1.11$\times$ for the three models compared to Megatron-LM's recommendations.

\subsection{Tuning Model Parallelism Configuration}\label{sec:tuner-parallel}
\begin{figure}[t]
	\centering
	\begin{subfigure}[t]{0.49\linewidth}
		\includegraphics[width=\linewidth]{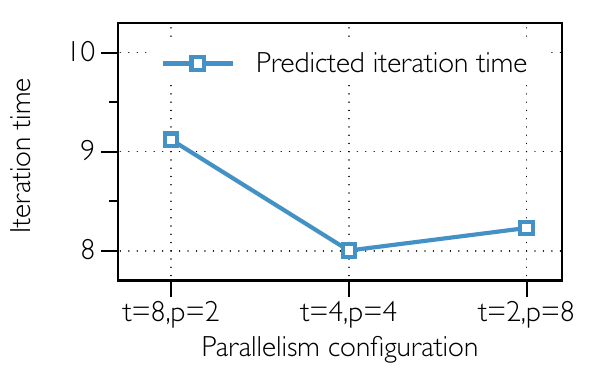}
		\caption{Predictions of \sys.}
		\label{fig:opti-parallel-1}
	\end{subfigure}
	\begin{subfigure}[t]{0.49\linewidth}
		\includegraphics[width=\linewidth]{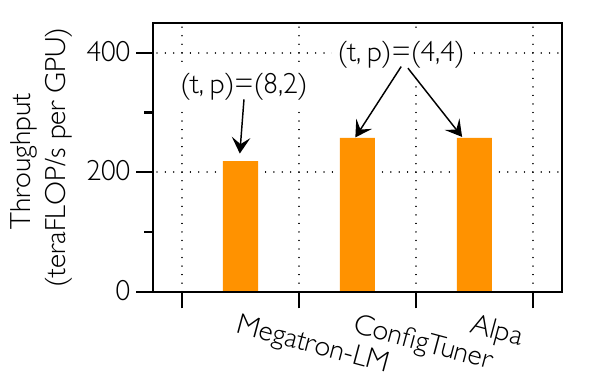}
		\caption{\sys vs. Megatron-LM vs. Alpa}
		\label{fig:opti-parallel-2}
	\end{subfigure}
	\caption{Tuning parallelism configs using \sys, achieving superior performance than Megatron-LM.}
	\label{fig:opti-parallel}
\end{figure}

For a given GPT model size, various model parallelism configurations are available for training. \sys can be used to predict the performance of different configurations and determine the most efficient one. For example, we evaluate a GPT model with 39B parameters, which can be distributed using multiple model parallelism configurations, such as $(t, p)$ of $(8,2)$, $(4,4)$, and $(2,8)$. Due to the high communication overhead in tensor parallelism, we restrict all tensor-parallel communications to be intra-node, while pipeline-parallel communications occur inter-node. The batch size is set to 48, and we test three configurations as inputs to \sys. For each configuration, we also vary the micro-batch size. \sys predicts the optimal micro-batch sizes for the $(8,2)$, $(4,4)$, and $(2,8)$ configurations to be 6, 3, and 1, respectively. Fig.~\ref{fig:opti-parallel-1} illustrates the predicted iteration time for three configurations under their optimal micro-batch sizes.

According to the results, \sys determines that the optimal parallelism configuration is $(4,4)$, which is the same as Alpa's suggestion. In contrast, the common configuration, which recommended by Megatron-LM, is $(8,2)$, following the principle that tensor parallelism should be maximized up to degree $g$ when using $g$-GPU servers~\cite{narayanan2021efficient}. We then conducted real-world experiments, and Fig.~\ref{fig:opti-parallel-2} presents the measured throughput for the configurations selected by \sys, Megatron-LM and Alpa, showing that \sys achieves a 1.17$\times$ improvement over Megatron-LM. 

Megatron-LM's recommendation is based on the assumption that intra-node interconnect capacity greatly exceeds inter-node capacity. However, our findings suggest that this generalized principle may not always be applicable, emphasizing the need for more precise analysis. \sys can serve as a valuable first-step analysis tool, with high accuracy and extremely low profiling overhead.

\subsection{Analyzing Data-Parallel Degree}\label{sec:tuner-dp}
Increasing the data-parallel (DP) degree is a widely used method to leverage additional computational resources and accelerate training. However, increasing the DP degree does not always result in proportional reductions in training time, and the associated costs can vary significantly. \sys can be used to analyze the trade-offs between training time reductions and costs across different DP degrees, helping identify the most cost-effective configurations.

\begin{figure}[t]
	\centering
	\begin{subfigure}[t]{0.495\linewidth}
		\includegraphics[width=\linewidth]{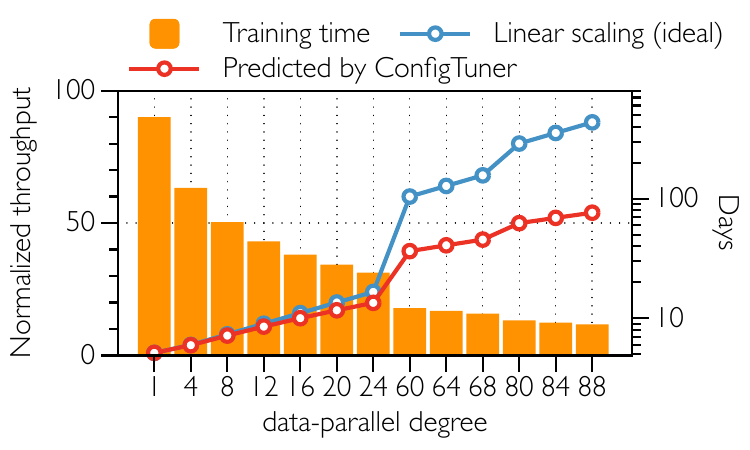}
		\caption{Predicted scaling factor and training time.}
		\label{fig:opti-dp-1}
	\end{subfigure}
	\begin{subfigure}[t]{0.495\linewidth}
		\includegraphics[width=\linewidth]{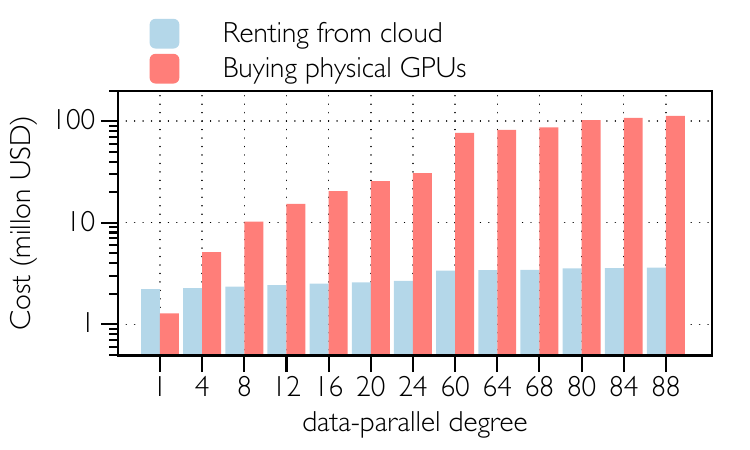}
		\caption{Predicted cost of renting and purchasing GPUs.}
		\label{fig:opti-dp-2}
	\end{subfigure}
	\caption{Using \sys to analyze the benefits and costs of increasing DP degree for a 145B GPT model.}
	\label{fig:opti-dp}
\end{figure}

For instance, we evaluate a 145B GPT model with a sequence length of 2048 and a global batch size of 2304. We vary the DP degree from 1 to 100 and adjust the corresponding batch size per model copy accordingly, while fixing the tensor- and pipeline-parallel degree to $(8,8)$. Using \sys, we predict the scaling factor and display it in Fig.~\ref{fig:opti-dp-1} (red line). For clarity, we demonstrate three representative ranges: (1) 1 to 24, (2) 60 to 68, and (3) 80 to 88. The ideal linear scaling is shown as a blue line for comparison. The results indicate that at lower DP degrees, such as in the 1 to 24 range, the scaling factor is nearly linear. However, as the DP degree exceeds 60, the scaling factor starts to flatten. Note that the flat scaling factor at high DP degrees is not caused by increased communication time, which stays nearly constant, but rather by the smaller batch size per model copy, which lowers GPU utilization and increases bubble time.

In addition to scaling factor, \sys can be used to evaluate the overall cost implications of increasing the DP degree. For example, when training a 145B model with 300B input tokens, \sys cam predict both the total training time and associated costs. Fig.~\ref{fig:opti-dp-1} shows the training time for this model at different DP degrees. In the range of 1 to 24, increasing the DP degree results in significant time savings, \eg, raising the DP degree from 4 to 8 reduces the training time from 123 days to 63 days. However, beyond DP degree 60, the benefits diminish considerably.

Fig.~\ref{fig:opti-dp-2} shows the corresponding costs for various DP degrees under two scenarios: renting GPUs from a cloud provider (assuming \$3 per GPU hour) and purchasing physical GPUs (assuming \$20,000 per H100 GPU). The analysis reveals that purchasing GPUs is cost-effective only when the DP degree is quite small. Below a certain degree, renting additional GPUs to increase the DP degree is a favorable option, as it significantly reduces training time without a substantial increase in costs. However, beyond this value, both renting and purchasing yield diminishing returns. For example, increasing the DP degree from 60 to 64 reduces the training time by just half a day (from 12 to 11.5 days) but incurs an additional cost of \$40,000 for renting and \$5 million for purchasing. Note that we cannot run Alpa for this analysis because conducting full iteration training at such a large scale (e.g., 3840 GPUs for DP degree 60) is too expensive. 


\section{Discussion}

\parab{Online profiling vs. mathematical analysis.} Many studies aim to enhance training performance through online profiling~\cite{xiao2018gandiva, rajasekaran2023cassini, sivathanu2019astra, wang2023topoopt, gu2019tiresias, hashemi2019tictac}, which relies on the repetitive feature observed across iterations. In contrast, our work moves beyond these understandings, finding that the communication can be accurately determined before execution (\ie, pre-execution accuracy). Furthermore, mathematical analysis has broader applications than online profiling. For example, while online profiling requires the realistic running of several training iteration, mathematical analysis can be applied to future or newly released GPUs to predict performance, as long as the GPU specifications are available. 


\parab{Extension to multi-job training scenario.} The goal of this paper is comprehensively analyzing predictability in single-job scenarios. However, it is feasible to extend our current analysis and cost model to multi-job scenarios~\cite{xiao2018gandiva, rajasekaran2023cassini, sivathanu2019astra}. The main difference in multi-job scenarios is the impact on effective bandwidth ($C$) due to link competition between jobs. For example, if two jobs share a link, the effective bandwidth typically becomes half of what it is when the link is used exclusively. Despite this, the core analysis remains the same as in single-job scenarios. We leave a rigorous analysis on multi-job training scenarios to our future work.

\section{Related Work}
\parab{Classical datacenter trace studies.} Numerous previous works have focused on statistical analyses of collected datacenter traces~\cite{benson2010understanding, roy2015inside, benson2010network, hu2024characterization, gangidi2024rdma, qian2024alibaba}. These studies typically conduct statistical analysis of large, mixed trace datasets to derive cluster-level metrics such as overall cluster utilization and hotspots for various job types. Due to the coarse-grained nature of this approach, prior studies on DNN training datacenter traces often only emphasize the basic repetitive feature~\cite{hu2024characterization, gangidi2024rdma, qian2024alibaba}. In contrast, our work studies the inherent features of DNN training systems and focuses on the fine-grained traffic characteristics inherent to these systems.

\parab{Extension to other training frameworks.} In addition to the hybrid parallel training, ZeRO~\cite{9355301} represents another distributed training system for LLMs. It partitions model states, including parameters, gradients, and optimizer state, across data-parallel processes, thereby eliminating memory redundancies. During training, these states are gathered as needed. Although our focus in this paper is on hybrid parallel training, our analytical methods, such as the analysis and dissection of iteration time, are also applicable to ZeRO, as ZeRO-based training exhibits similar predictability characteristics.

\parab{Comparision with automatic parallelism searching.} Many prior works, such as FlexFlow~\cite{jia2019beyond} and Alpa~\cite{zheng2022alpa}, focus on automatically exploring the parallelism strategy search space to find the most efficient configurations. As shown in $\S$\ref{optimization}, our analytical formulation could be used to optimize parallelism configurations with extremely low searching complexity. Despite this, the primary aim of this paper is orthogonal to these automatic parallelism search approaches. Our goal is to deepen the understanding of predictability and provide insightful heuristics, rather than developing general-purpose parallelism space exploration algorithms. 





\section{Conclusion} 
In this paper, we systematically formulate and experimentally validate the predictability of LLM training. We believe our rigorous modeling and comprehensive analysis provides a deeper understanding for researchers in ML systems and offers new insights into training optimization.




\bibliographystyle{plain}
\bibliography{main.bib}

@article{mccandlish2018empirical,
  title={An empirical model of large-batch training},
  author={McCandlish, Sam and Kaplan, Jared and Amodei, Dario and Team, OpenAI Dota},
  journal={arXiv preprint arXiv:1812.06162},
  year={2018}
}

@inproceedings{qian2024alibaba,
  title={Alibaba HPN: a data center network for large language model training},
  author={Qian, Kun and Xi, Yongqing and Cao, Jiamin and Gao, Jiaqi and Xu, Yichi and Guan, Yu and Fu, Binzhang and Shi, Xuemei and Zhu, Fangbo and Miao, Rui and others},
  booktitle={Proceedings of the ACM SIGCOMM 2024 Conference},
  pages={691--706},
  year={2024}
}

@inproceedings{gangidi2024rdma,
  title={RDMA over Ethernet for Distributed Training at Meta Scale},
  author={Gangidi, Adithya and Miao, Rui and Zheng, Shengbao and Bondu, Sai Jayesh and Goes, Guilherme and Morsy, Hany and Puri, Rohit and Riftadi, Mohammad and Shetty, Ashmitha Jeevaraj and Yang, Jingyi and others},
  booktitle={Proceedings of the ACM SIGCOMM 2024 Conference},
  pages={57--70},
  year={2024}
}

@article{benson2010understanding,
  title={Understanding data center traffic characteristics},
  author={Benson, Theophilus and Anand, Ashok and Akella, Aditya and Zhang, Ming},
  journal={ACM SIGCOMM Computer Communication Review},
  volume={40},
  number={1},
  pages={92--99},
  year={2010},
  publisher={ACM New York, NY, USA}
}

@inproceedings{roy2015inside,
  title={Inside the social network's (datacenter) network},
  author={Roy, Arjun and Zeng, Hongyi and Bagga, Jasmeet and Porter, George and Snoeren, Alex C},
  booktitle={Proceedings of the 2015 ACM Conference on Special Interest Group on Data Communication},
  pages={123--137},
  year={2015}
}

@inproceedings{benson2010network,
  title={Network traffic characteristics of data centers in the wild},
  author={Benson, Theophilus and Akella, Aditya and Maltz, David A},
  booktitle={Proceedings of the 10th ACM SIGCOMM conference on Internet measurement},
  pages={267--280},
  year={2010}
}

@article{yang2024qwen2,
  title={Qwen2 technical report},
  author={Yang, An and Yang, Baosong and Hui, Binyuan and Zheng, Bo and Yu, Bowen and Zhou, Chang and Li, Chengpeng and Li, Chengyuan and Liu, Dayiheng and Huang, Fei and others},
  journal={arXiv preprint arXiv:2407.10671},
  year={2024}
}

@article{dai2024deepseekmoe,
  title={Deepseekmoe: Towards ultimate expert specialization in mixture-of-experts language models},
  author={Dai, Damai and Deng, Chengqi and Zhao, Chenggang and Xu, RX and Gao, Huazuo and Chen, Deli and Li, Jiashi and Zeng, Wangding and Yu, Xingkai and Wu, Y and others},
  journal={arXiv preprint arXiv:2401.06066},
  year={2024}
}

@misc{superpod,
  author = "NVIDIA DGX SuperPOD Infrastructure",
  howpublished = "https://www.nvidia.com/en-us/data-center/dgx-superpod/",
  year = "2024"}

@misc{nccl,
  author = "NCCL",
  howpublished = "https://github.com/NVIDIA/nccl",
  year = "2024"}

@misc{ds,
	author = "Megatron-DeepSpeed",
	howpublished = "https://github.com/microsoft/Megatron-DeepSpeed",
	year="2021"}

@article{lepikhin2020gshard,
  title={Gshard: Scaling giant models with conditional computation and automatic sharding},
  author={Lepikhin, Dmitry and Lee, HyoukJoong and Xu, Yuanzhong and Chen, Dehao and Firat, Orhan and Huang, Yanping and Krikun, Maxim and Shazeer, Noam and Chen, Zhifeng},
  journal={arXiv preprint arXiv:2006.16668},
  year={2020}
}

@inproceedings{rajbhandari2022deepspeed,
  title={Deepspeed-moe: Advancing mixture-of-experts inference and training to power next-generation ai scale},
  author={Rajbhandari, Samyam and Li, Conglong and Yao, Zhewei and Zhang, Minjia and Aminabadi, Reza Yazdani and Awan, Ammar Ahmad and Rasley, Jeff and He, Yuxiong},
  booktitle={International conference on machine learning},
  pages={18332--18346},
  year={2022},
  organization={PMLR}
}

@INPROCEEDINGS{9355301,
  author={Rajbhandari, Samyam and Rasley, Jeff and Ruwase, Olatunji and He, Yuxiong},
  booktitle={SC20: International Conference for High Performance Computing, Networking, Storage and Analysis}, 
  title={ZeRO: Memory optimizations Toward Training Trillion Parameter Models}, 
  year={2020},
  volume={},
  number={},
  pages={1-16},
  doi={10.1109/SC41405.2020.00024}}

@article{zhao2023survey,
  title={A survey of large language models},
  author={Zhao, Wayne Xin and Zhou, Kun and Li, Junyi and Tang, Tianyi and Wang, Xiaolei and Hou, Yupeng and Min, Yingqian and Zhang, Beichen and Zhang, Junjie and Dong, Zican and others},
  journal={arXiv preprint arXiv:2303.18223},
  year={2023}
}

@inproceedings{zhu2020daydream,
  title={Daydream: Accurately estimating the efficacy of optimizations for $\{$DNN$\}$ training},
  author={Zhu, Hongyu and Phanishayee, Amar and Pekhimenko, Gennady},
  booktitle={2020 USENIX Annual Technical Conference (USENIX ATC 20)},
  pages={337--352},
  year={2020}
}

@inproceedings{liu2022modeling,
  title={Modeling and optimizing the scaling performance in distributed deep learning training},
  author={Liu, Ting and Miao, Tianhao and Wu, Qinghua and Li, Zhenyu and He, Guangxin and Wu, Jiaoren and Zhang, Shengzhuo and Yang, Xingwu and Tyson, Gareth and Xie, Gaogang},
  booktitle={Proceedings of the ACM Web Conference 2022},
  pages={1764--1773},
  year={2022}
}

@inproceedings{sivathanu2019astra,
  title={Astra: Exploiting predictability to optimize deep learning},
  author={Sivathanu, Muthian and Chugh, Tapan and Singapuram, Sanjay S and Zhou, Lidong},
  booktitle={Proceedings of the Twenty-Fourth International Conference on Architectural Support for Programming Languages and Operating Systems},
  pages={909--923},
  year={2019}
}

@misc{wang2023build,
      title={How to Build Low-cost Networks for Large Language Models (without Sacrificing Performance)?}, 
      author={Weiyang Wang and Manya Ghobadi and Kayvon Shakeri and Ying Zhang and Naader Hasani},
      year={2023},
      eprint={2307.12169},
      archivePrefix={arXiv},
      primaryClass={cs.NI}
}

@inproceedings{wang2023topoopt,
  title={$\{$TopoOpt$\}$: Co-optimizing Network Topology and Parallelization Strategy for Distributed Training Jobs},
  author={Wang, Weiyang and Khazraee, Moein and Zhong, Zhizhen and Ghobadi, Manya and Jia, Zhihao and Mudigere, Dheevatsa and Zhang, Ying and Kewitsch, Anthony},
  booktitle={20th USENIX Symposium on Networked Systems Design and Implementation (NSDI 23)},
  pages={739--767},
  year={2023}
}

@inproceedings{xiao2018gandiva,
  title={Gandiva: Introspective cluster scheduling for deep learning},
  author={Xiao, Wencong and Bhardwaj, Romil and Ramjee, Ramachandran and Sivathanu, Muthian and Kwatra, Nipun and Han, Zhenhua and Patel, Pratyush and Peng, Xuan and Zhao, Hanyu and Zhang, Quanlu and others},
  booktitle={13th USENIX Symposium on Operating Systems Design and Implementation (OSDI 18)},
  pages={595--610},
  year={2018}
}

@article{rajasekaran2023cassini,
  title={Cassini: Network-Aware Job Scheduling in Machine Learning Clusters},
  author={Rajasekaran, Sudarsanan and Ghobadi, Manya and Akella, Aditya},
  journal={arXiv preprint arXiv:2308.00852},
  year={2023}
}

@article{jayarajan2019priority,
  title={Priority-based parameter propagation for distributed DNN training},
  author={Jayarajan, Anand and Wei, Jinliang and Gibson, Garth and Fedorova, Alexandra and Pekhimenko, Gennady},
  journal={Proceedings of Machine Learning and Systems},
  volume={1},
  pages={132--145},
  year={2019}
}

@inproceedings{ma2022autobyte,
  title={Autobyte: Automatic configuration for optimal communication scheduling in dnn training},
  author={Ma, Yiqing and Wang, Hao and Zhang, Yiming and Chen, Kai},
  booktitle={IEEE INFOCOM 2022-IEEE Conference on Computer Communications},
  pages={760--769},
  year={2022},
  organization={IEEE}
}

@inproceedings{fei2021efficient,
  title={Efficient sparse collective communication and its application to accelerate distributed deep learning},
  author={Fei, Jiawei and Ho, Chen-Yu and Sahu, Atal N and Canini, Marco and Sapio, Amedeo},
  booktitle={Proceedings of the 2021 ACM SIGCOMM 2021 Conference},
  pages={676--691},
  year={2021}
}

@inproceedings{li2024thc,
  title={$\{$THC$\}$: Accelerating Distributed Deep Learning Using Tensor Homomorphic Compression},
  author={Li, Minghao and Basat, Ran Ben and Vargaftik, Shay and Lao, ChonLam and Xu, Kevin and Mitzenmacher, Michael and Yu, Minlan},
  booktitle={21st USENIX Symposium on Networked Systems Design and Implementation (NSDI 24)},
  pages={1191--1211},
  year={2024}
}

@inproceedings{bai2021gradient,
  title={Gradient compression supercharged high-performance data parallel dnn training},
  author={Bai, Youhui and Li, Cheng and Zhou, Quan and Yi, Jun and Gong, Ping and Yan, Feng and Chen, Ruichuan and Xu, Yinlong},
  booktitle={Proceedings of the ACM SIGOPS 28th Symposium on Operating Systems Principles},
  pages={359--375},
  year={2021}
}

@inproceedings{wang2023hi,
  title={Hi-Speed DNN Training with Espresso: Unleashing the Full Potential of Gradient Compression with Near-Optimal Usage Strategies},
  author={Wang, Zhuang and Lin, Haibin and Zhu, Yibo and Ng, TS Eugene},
  booktitle={Proceedings of the Eighteenth European Conference on Computer Systems},
  pages={867--882},
  year={2023}
}

@article{huang2019gpipe,
  title={Gpipe: Efficient training of giant neural networks using pipeline parallelism},
  author={Huang, Yanping and Cheng, Youlong and Bapna, Ankur and Firat, Orhan and Chen, Dehao and Chen, Mia and Lee, HyoukJoong and Ngiam, Jiquan and Le, Quoc V and Wu, Yonghui and others},
  journal={Advances in neural information processing systems},
  volume={32},
  year={2019}
}

@article{cho2019blueconnect,
  title={Blueconnect: Decomposing all-reduce for deep learning on heterogeneous network hierarchy},
  author={Cho, Minsik and Finkler, Ulrich and Kung, David and Hunter, Hillery},
  journal={Proceedings of Machine Learning and Systems},
  volume={1},
  pages={241--251},
  year={2019}
}

@article{thakur2005optimization,
  title={Optimization of collective communication operations in MPICH},
  author={Thakur, Rajeev and Rabenseifner, Rolf and Gropp, William},
  journal={The International Journal of High Performance Computing Applications},
  volume={19},
  number={1},
  pages={49--66},
  year={2005},
  publisher={Sage Publications Sage CA: Thousand Oaks, CA}
}

@article{hu2024characterization,
  title={Characterization of large language model development in the datacenter},
  author={Hu, Qinghao and Ye, Zhisheng and Wang, Zerui and Wang, Guoteng and Zhang, Meng and Chen, Qiaoling and Sun, Peng and Lin, Dahua and Wang, Xiaolin and Luo, Yingwei and others},
  journal={arXiv preprint arXiv:2403.07648},
  year={2024}
}

@inproceedings{jiang2024megascale,
  title={$\{$MegaScale$\}$: Scaling Large Language Model Training to More Than 10,000 $\{$GPUs$\}$},
  author={Jiang, Ziheng and Lin, Haibin and Zhong, Yinmin and Huang, Qi and Chen, Yangrui and Zhang, Zhi and Peng, Yanghua and Li, Xiang and Xie, Cong and Nong, Shibiao and others},
  booktitle={21st USENIX Symposium on Networked Systems Design and Implementation (NSDI 24)},
  pages={745--760},
  year={2024}
}

@inproceedings{zeng2024accelerating,
  title={Accelerating Neural Recommendation Training with Embedding Scheduling},
  author={Zeng, Chaoliang and Liao, Xudong and Cheng, Xiaodian and Tian, Han and Wan, Xinchen and Wang, Hao and Chen, Kai},
  booktitle={21st USENIX Symposium on Networked Systems Design and Implementation (NSDI 24)},
  pages={1141--1156},
  year={2024}
}

@article{jia2019beyond,
  title={Beyond Data and Model Parallelism for Deep Neural Networks.},
  author={Jia, Zhihao and Zaharia, Matei and Aiken, Alex},
  journal={Proceedings of Machine Learning and Systems},
  volume={1},
  pages={1--13},
  year={2019}
}

@inproceedings{zheng2022alpa,
  title={Alpa: Automating inter-and $\{$Intra-Operator$\}$ parallelism for distributed deep learning},
  author={Zheng, Lianmin and Li, Zhuohan and Zhang, Hao and Zhuang, Yonghao and Chen, Zhifeng and Huang, Yanping and Wang, Yida and Xu, Yuanzhong and Zhuo, Danyang and Xing, Eric P and others},
  booktitle={16th USENIX Symposium on Operating Systems Design and Implementation (OSDI 22)},
  pages={559--578},
  year={2022}
}

@inproceedings{peng2019generic,
  title={A generic communication scheduler for distributed DNN training acceleration},
  author={Peng, Yanghua and Zhu, Yibo and Chen, Yangrui and Bao, Yixin and Yi, Bairen and Lan, Chang and Wu, Chuan and Guo, Chuanxiong},
  booktitle={Proceedings of the 27th ACM Symposium on Operating Systems Principles},
  pages={16--29},
  year={2019}
}

@article{brown2020language,
  title={Language models are few-shot learners},
  author={Brown, Tom and Mann, Benjamin and Ryder, Nick and Subbiah, Melanie and Kaplan, Jared D and Dhariwal, Prafulla and Neelakantan, Arvind and Shyam, Pranav and Sastry, Girish and Askell, Amanda and others},
  journal={Advances in neural information processing systems},
  volume={33},
  pages={1877--1901},
  year={2020}
}

@article{touvron2023llama,
  title={Llama: Open and efficient foundation language models},
  author={Touvron, Hugo and Lavril, Thibaut and Izacard, Gautier and Martinet, Xavier and Lachaux, Marie-Anne and Lacroix, Timoth{\'e}e and Rozi{\`e}re, Baptiste and Goyal, Naman and Hambro, Eric and Azhar, Faisal and others},
  journal={arXiv preprint arXiv:2302.13971},
  year={2023}
}

@article{fedus2022switch,
  title={Switch transformers: Scaling to trillion parameter models with simple and efficient sparsity},
  author={Fedus, William and Zoph, Barret and Shazeer, Noam},
  journal={Journal of Machine Learning Research},
  volume={23},
  number={120},
  pages={1--39},
  year={2022}
}

@article{shazeer2017outrageously,
  title={Outrageously large neural networks: The sparsely-gated mixture-of-experts layer},
  author={Shazeer, Noam and Mirhoseini, Azalia and Maziarz, Krzysztof and Davis, Andy and Le, Quoc and Hinton, Geoffrey and Dean, Jeff},
  journal={arXiv preprint arXiv:1701.06538},
  year={2017}
}

@article{wang2020domain,
  title={Domain-specific communication optimization for distributed DNN training},
  author={Wang, Hao and Chen, Jingrong and Wan, Xinchen and Tian, Han and Xia, Jiacheng and Zeng, Gaoxiong and Wang, Weiyan and Chen, Kai and Bai, Wei and Jiang, Junchen},
  journal={arXiv preprint arXiv:2008.08445},
  year={2020}
}

@inproceedings{jiang2020unified,
  title={A unified architecture for accelerating distributed $\{$DNN$\}$ training in heterogeneous $\{$GPU/CPU$\}$ clusters},
  author={Jiang, Yimin and Zhu, Yibo and Lan, Chang and Yi, Bairen and Cui, Yong and Guo, Chuanxiong},
  booktitle={14th USENIX Symposium on Operating Systems Design and Implementation (OSDI 20)},
  pages={463--479},
  year={2020}
}

@article{li2014communication,
  title={Communication efficient distributed machine learning with the parameter server},
  author={Li, Mu and Andersen, David G and Smola, Alexander J and Yu, Kai},
  journal={Advances in Neural Information Processing Systems},
  volume={27},
  year={2014}
}

@inproceedings{narayanan2019pipedream,
  title={PipeDream: Generalized pipeline parallelism for DNN training},
  author={Narayanan, Deepak and Harlap, Aaron and Phanishayee, Amar and Seshadri, Vivek and Devanur, Nikhil R and Ganger, Gregory R and Gibbons, Phillip B and Zaharia, Matei},
  booktitle={Proceedings of the 27th ACM Symposium on Operating Systems Principles},
  pages={1--15},
  year={2019}
}

@article{korthikanti2023reducing,
  title={Reducing activation recomputation in large transformer models},
  author={Korthikanti, Vijay Anand and Casper, Jared and Lym, Sangkug and McAfee, Lawrence and Andersch, Michael and Shoeybi, Mohammad and Catanzaro, Bryan},
  journal={Proceedings of Machine Learning and Systems},
  volume={5},
  year={2023}
}

@inproceedings{narayanan2021efficient,
  title={Efficient large-scale language model training on gpu clusters using megatron-lm},
  author={Narayanan, Deepak and Shoeybi, Mohammad and Casper, Jared and LeGresley, Patrick and Patwary, Mostofa and Korthikanti, Vijay and Vainbrand, Dmitri and Kashinkunti, Prethvi and Bernauer, Julie and Catanzaro, Bryan and others},
  booktitle={Proceedings of the International Conference for High Performance Computing, Networking, Storage and Analysis},
  pages={1--15},
  year={2021}
}

@article{radford2019language,
  title={Language models are unsupervised multitask learners},
  author={Radford, Alec and Wu, Jeffrey and Child, Rewon and Luan, David and Amodei, Dario and Sutskever, Ilya and others},
  journal={OpenAI blog},
  volume={1},
  number={8},
  pages={9},
  year={2019}
}

@inproceedings{zhang2020network,
  title={Is network the bottleneck of distributed training?},
  author={Zhang, Zhen and Chang, Chaokun and Lin, Haibin and Wang, Yida and Arora, Raman and Jin, Xin},
  booktitle={Proceedings of the Workshop on Network Meets AI \& ML},
  pages={8--13},
  year={2020}
}

@inproceedings{gu2019tiresias,
  title={Tiresias: A $\{$GPU$\}$ cluster manager for distributed deep learning},
  author={Gu, Juncheng and Chowdhury, Mosharaf and Shin, Kang G and Zhu, Yibo and Jeon, Myeongjae and Qian, Junjie and Liu, Hongqiang and Guo, Chuanxiong},
  booktitle={16th USENIX Symposium on Networked Systems Design and Implementation (NSDI 19)},
  pages={485--500},
  year={2019}
}

@article{hashemi2019tictac,
  title={Tictac: Accelerating distributed deep learning with communication scheduling},
  author={Hashemi, Sayed Hadi and Abdu Jyothi, Sangeetha and Campbell, Roy},
  journal={Proceedings of Machine Learning and Systems},
  volume={1},
  pages={418--430},
  year={2019}
}

\appendix
\section{Intra-node GPU interconnect topologies}\label{apx:topo}
\begin{figure}[h]
	\centering
	\includegraphics[width=\linewidth]{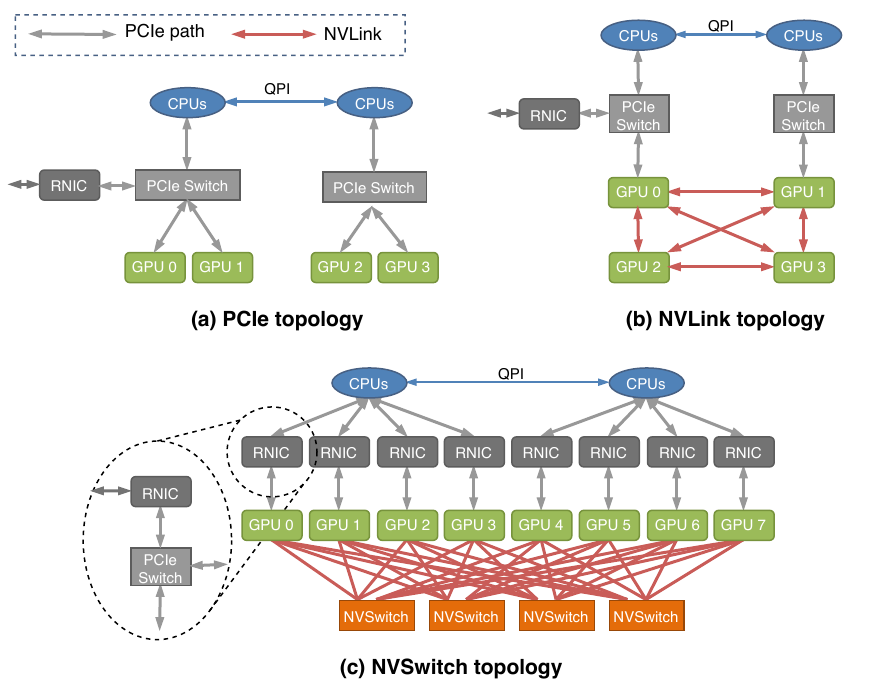}
	\caption{Illustration of typical intra-node GPU interconnect topologies: PCIe, NVLink and NVSwitch. A single red line represents two NVLink lines.}
	\label{fig:interconnect}
\end{figure}

The three GPU interconnect topologies of PCIe, NVLink and NVSwitch are illustrated in Fig.~\ref{fig:interconnect}.

\textit{(1) PCIe}: Fig.~\ref{fig:interconnect}\violet{a} depicts the PCIe topology, which forms a balanced tree structure. Several GPUs are interconnected via a PCIe switch, which is then connected to a CPU host bridge. A similar arrangement applies to other GPUs. Finally, the CPU host bridges are linked via QuickPath Interconnect (QPI). For GPU-to-RNIC connectivity, each machine is equipped with a single RNIC that connects to the PCIe switch.

\textit{(2) NVLink}: Fig.~\ref{fig:interconnect}\violet{b} displays the NVLink topology used in our Tesla-NVLink platform. Each GPU features 6 NVLink slots, each offering 25GB/s (unidirectional) bandwidth per link. The four GPUs form a fully-connected topology, meaning any two GPUs can communicate directly. The GPUs are also connected via a PCIe network. Similar to the PCIe topology, each machine has one RNIC linked to a PCIe switch.

\textit{(3) NVSwitch}: Fig.~\ref{fig:interconnect}\violet{c} presents the NVSwitch topology used in our Hopper platform. Each GPU features 8 NVLink slots, each offering 25GB/s (unidirectional) bandwidth. These GPUs connect to 4 NVSwitches, with 2 NVLink connections per NVSwitch. For GPU-to-RNIC connectivity, the Hopper platform includes 8 RNICs per machine, each connected to a different PCIe switch.

\end{document}